\documentclass[aps,prl,preprint,amsmath,amssymb,showpacs,superscriptaddress]{revtex4-2}

\usepackage{graphicx}
\usepackage{dcolumn}
\usepackage{bm}
\usepackage{graphicx}
\usepackage{overpic}
\usepackage{longtable}
\usepackage{epsfig}
\usepackage{epsf}
\usepackage{booktabs}
\usepackage{colortbl}
\usepackage{booktabs}
\usepackage{amssymb}
\usepackage{amsmath}
\usepackage{amsthm}
\usepackage{mathtools}
\usepackage{braket}
\usepackage{multirow}
\usepackage{verbatim}
\usepackage[T1]{fontenc}
\usepackage{amsfonts}
\usepackage{scalerel}
\usepackage{cases}
\usepackage{pgfplots}
\usepackage{float}
\usepackage{lipsum}
\usepackage{gensymb}
\usepackage[colorlinks=true,linkcolor=blue,citecolor=blue]{hyperref}

\newcommand{\myfont}{\fontfamily{phv}\selectfont}
\newcommand{\figfont}{\myfont \textsf \large}

\makeatletter
\begin{document}

\title{Molecular chiral discrimination through symmetry-breaking spin dynamics}

\author{Tianyu Xie}
\thanks{T. Xie and Y. Hao contributed equally to this work.}
\affiliation{Laboratory of Spin Magnetic Resonance, School of Physical Sciences, Anhui Province Key Laboratory of Scientific Instrument Development and Application, University of Science and Technology of China, Hefei 230026, China}
\affiliation{Hefei National Research Center for Physical Sciences at the Microscale, University of Science and Technology of China, Hefei 230026, China}

\author{Yucheng Hao}
\thanks{T. Xie and Y. Hao contributed equally to this work.}
\affiliation{Laboratory of Spin Magnetic Resonance, School of Physical Sciences, Anhui Province Key Laboratory of Scientific Instrument Development and Application, University of Science and Technology of China, Hefei 230026, China}
\affiliation{Hefei National Research Center for Physical Sciences at the Microscale, University of Science and Technology of China, Hefei 230026, China}

\author{Mingzhe Liu}
\affiliation{Laboratory of Spin Magnetic Resonance, School of Physical Sciences, Anhui Province Key Laboratory of Scientific Instrument Development and Application, University of Science and Technology of China, Hefei 230026, China}

\author{Mengqi Wang}
\affiliation{Laboratory of Spin Magnetic Resonance, School of Physical Sciences, Anhui Province Key Laboratory of Scientific Instrument Development and Application, University of Science and Technology of China, Hefei 230026, China}
\affiliation{Hefei National Research Center for Physical Sciences at the Microscale, University of Science and Technology of China, Hefei 230026, China}

\author{Shaoyi Xu}
\affiliation{Laboratory of Spin Magnetic Resonance, School of Physical Sciences, Anhui Province Key Laboratory of Scientific Instrument Development and Application, University of Science and Technology of China, Hefei 230026, China}

\author{Zhiyuan Zhao}
\affiliation{Laboratory of Spin Magnetic Resonance, School of Physical Sciences, Anhui Province Key Laboratory of Scientific Instrument Development and Application, University of Science and Technology of China, Hefei 230026, China}
\affiliation{Hefei National Research Center for Physical Sciences at the Microscale, University of Science and Technology of China, Hefei 230026, China}

\author{Chang-Kui Duan}
\affiliation{Laboratory of Spin Magnetic Resonance, School of Physical Sciences, Anhui Province Key Laboratory of Scientific Instrument Development and Application, University of Science and Technology of China, Hefei 230026, China}
\affiliation{Hefei National Research Center for Physical Sciences at the Microscale, University of Science and Technology of China, Hefei 230026, China}
\affiliation{Hefei National Laboratory, University of Science and Technology of China, Hefei 230088, China}

\author{Ya Wang}
\affiliation{Laboratory of Spin Magnetic Resonance, School of Physical Sciences, Anhui Province Key Laboratory of Scientific Instrument Development and Application, University of Science and Technology of China, Hefei 230026, China}
\affiliation{Hefei National Research Center for Physical Sciences at the Microscale, University of Science and Technology of China, Hefei 230026, China}
\affiliation{Hefei National Laboratory, University of Science and Technology of China, Hefei 230088, China}

\author{Fazhan Shi}
\email{fzshi@ustc.edu.cn}
\affiliation{Laboratory of Spin Magnetic Resonance, School of Physical Sciences, Anhui Province Key Laboratory of Scientific Instrument Development and Application, University of Science and Technology of China, Hefei 230026, China}
\affiliation{Hefei National Research Center for Physical Sciences at the Microscale, University of Science and Technology of China, Hefei 230026, China}
\affiliation{Hefei National Laboratory, University of Science and Technology of China, Hefei 230088, China}
\affiliation{School of Biomedical Engineering and Suzhou Institute for Advanced Research, University of Science and Technology of China, Suzhou 215123, China}
\affiliation{The First Affiliated Hospital of USTC, Division of Life Sciences and Medicine, University of Science and Technology of China, Hefei 230026, China}

\author{Jiangfeng Du}
\affiliation{State Key Laboratory of Ocean Sensing and School of Physics, Zhejiang University, Hangzhou 310058, China}

\begin{abstract}
Molecular chirality plays a crucial role in physics, chemistry, life sciences and pharmacology. Nowadays, the chiral discrimination and control at the single-molecule level is urgently needed to reveal the origin of the chirality-relevant phenomena by recovering the information disturbed by the ensemble averaging. The method of magnetic resonance (MR), as one of powerful tools for structure analysis, is blind to the molecular chirality in the absence of a chiral reagent. Here we propose and experimentally demonstrate a direct MR-based method for determining the chirality at the single-molecule level through constructing the symmetry-breaking dynamics of nearby nuclear spins. In principle, the mirror asymmetry of two enantiomers in real space is manifested by breaking the joint symmetry of the mirror reflection and time reversal in spin space under spin dynamics. Experimentally, two enantiomers are indistinguishable from the dynamics of strongly-coupled but unpolarized nuclear spins, but diverge evidently in the dynamical results that break the field-inversion symmetry after spins are polarized. Our method and results will benefit the study of chirality-induced properties in the fields of chemistry and biology.
\end{abstract}

\maketitle

Since the seminal work of Pasteur on the molecular chirality~\cite{pasteur1848recherches}, numerous chirality-related phenomena have been discovered and studied extensively in modern science. Among them, the most prominent are parity non-conservation in weak interactions~\cite{lee1956question,wu1957experimental}, catalytic asymmetric synthesis~\cite{akiyama2022catalytic}, biological homochirality~\cite{guijarro2008origin}, and chiral pharmaceuticals~\cite{nguyen2006chiral}. Many recent developments in this field, such as the computer-assisted design~\cite{zahrt2019quantitative,pinus2024computational} for enantioselective catalysis~\cite{brimioulle2015enantioselective,zi2016recent,mondal2022enantioselective}, the effect of chiral-induced spin selectivity (CISS)~\cite{ray1999asymmetric,gohler2011spin,bloom2024chiral}, chiral nanostructures and supramolecular assemblies~\cite{sun2022chirality,hembury2008chirality,liu2015supramolecular} and so on, demand more in-depth investigations at the microscale, even at the single-molecule resolution, to elucidate the mechanisms underlying these chiral-induced properties or processes.

Conventional magnetic resonance (MR) spectroscopy is a non-destructive and versatile analytical method, and thus seems to be an ideal platform for chiral discrimination. However, MR-based methods are supposed to be unable to directly distinguish chemical enantiomers. The MR methods through detecting the small energy shifts induced by parity-violating weak interactions~\cite{bast2011analysis,eills2017measuring,quack2022perspectives} or the pseudoscalar terms by applying electric fields~\cite{buckingham2004chirality,buckingham2006direct,walls2008measuring} are still unavailable. To destroy the enantiomeric symmetry, chiral reagents, like chiral derivatizing agents, chiral solvents or metal complexes, have to be added before the MR detection in the previous experiments~\cite{wenzel2011using,silva2017recent}. Such indirect MR-based methods are quite cumbersome, compared with non-MR methods like chiral chromatography~\cite{ward2012chiral}, circular dichroism spectroscopy~\cite{ranjbar2009circular} or X-ray crystallography~\cite{flack2008use}. Moreover, these mainstream approaches for chiral discrimination, including MR-based ones, are not sensitive enough to detect the molecular chirality at the single-molecule level.


In this study, we propose an MR-based method to directly distinguish the chirality through constructing and probing symmetry-breaking dynamics of the nuclear spins near the chiral center, and the principle of this method is elucidated from the perspective of symmetry breaking. Combined with the single-molecule MR technology~\cite{shi2015single,lovchinsky2016nuclear,lovchinsky2017magnetic,shi2018single,du2024single} using the nitrogen-vacancy (NV) center in diamond, our method further enables the investigation of chiral molecules at the single-molecule resolution. The experiments in this work are performed on single pseudomolecules comprised of the NV center and the nearby carbon atoms. As proof of principle, the configurations of two enantiomers are correlated to the dynamics of two nuclear spins that are partially polarized and J-coupled. Moreover, it is observed after reversing the magnetic field that the dynamics corresponding to two enantiomers are mutually exchanged, as predicted by the pseudoscalar. The chirality can be also correlated to the dynamics of three or more nuclear spins with three-body dynamics recorded to determine the chirality. Finally, the symmetry breaking process is explicitly exhibited as a function of the polarization, concluding that direct chiral discrimination is achieved by breaking the mirror reflection and time reversal symmetry (MT-symmetry) in spin dynamics.

\begin{figure}
\centering
\begin{overpic}[width=1.0\columnwidth]{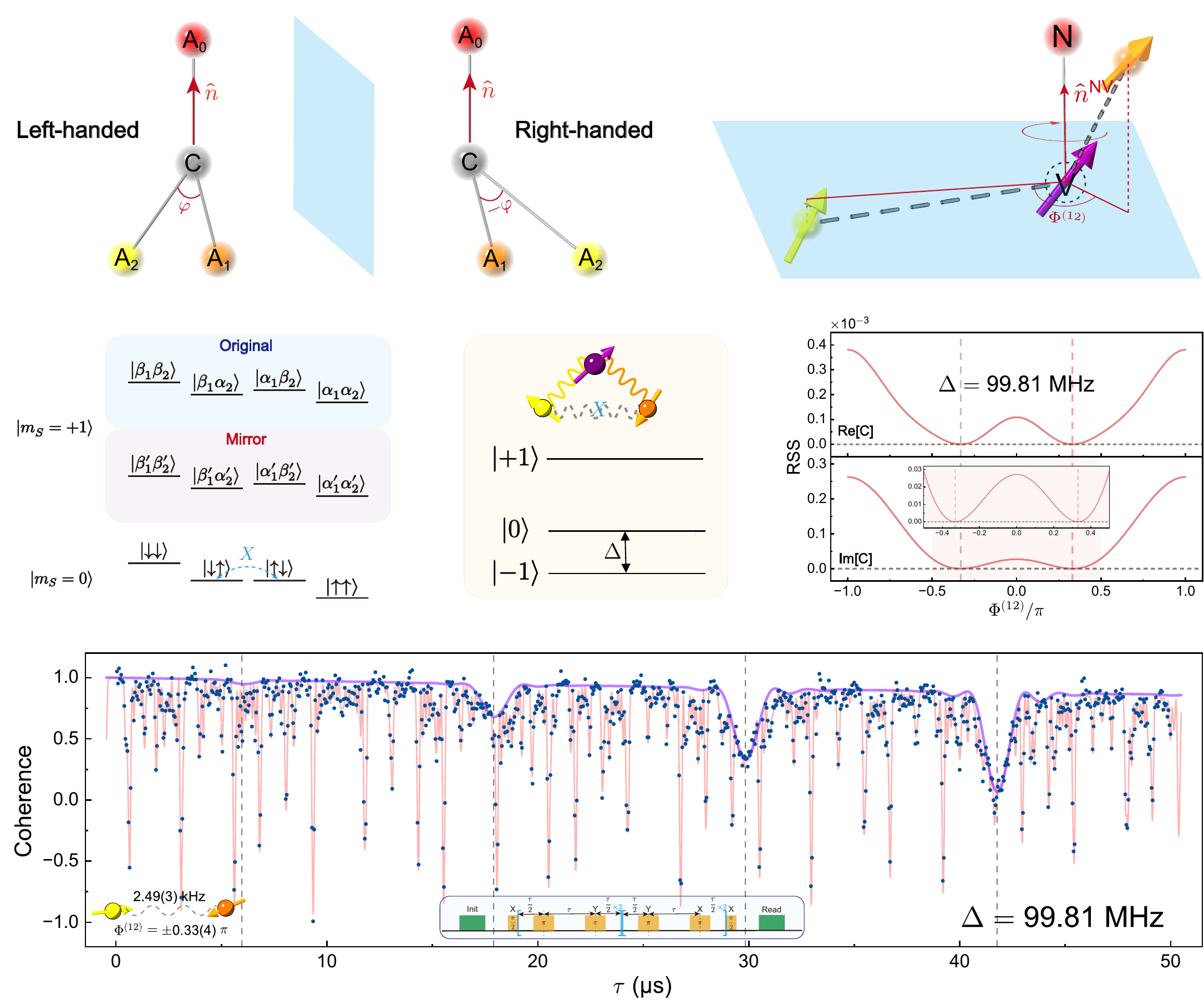}
	\put (0.4, 81.3) {\figfont{(a)}}
	\put (56., 81.3) {\figfont{(b)}}
	\put (0.4, 55.5) {\figfont{(c)}}
	\put (35.9, 55.5) {\figfont{(d)}}
	\put (0.4, 28.9) {\figfont{(e)}}
	\put (64.5, 55.5) {\figfont{(f)}}
\end{overpic}
\caption{Chiral molecules, chiral discrimination and dynamics of the nearby nuclear spins. (a) Simplified schematic of two mirror-symmetric enantiomers of a chiral molecule, labeled by the sign of the angle between CA$_{\text{1}}$ and CA$_{\text{2}}$. (b) The molecular system used for detecting the chirality indicated by the angle $\Phi^{(12)}$. (c) Level structure of the NV electron spin and two $^{\text{13}}$C nuclear spins. Two enantiomers have identical eigenenergies, but differ in the eigenstates of two $^{\text{13}}$C spins in the subspace $\ket{m_S=+1}$. (d) Induced J-coupling between two $^{\text{13}}$C spins by tuning the energy gap $\Delta$ of the NV spin. (e) The dynamical results of two unpolarized $^{\text{13}}$C spins measured under $\Delta$ = 99.81~MHz by applying the XY8 sequence. The red line is the fitting curve with $\Phi^{(12)}$ = $\pm$0.33(4)~$\pi$ and the purple one depicts the coupling-induced envelope. (f) The difference between the spin coherence as a function of the phase $\Phi^{(12)}$ and the fitting curve in (e). RSS, short for the residual squared sum, gives a metric of the difference, and Re[C] (Im[C]) means the real (imaginary) part of the coherence.}\label{chiral discrimination}
\end{figure}

Chirality, or handedness, refers to a three-dimensional asymmetry that an object is not superimposable on its mirror image, as shown in Fig.~\ref{chiral discrimination}(a). In general, there always exists a pseudoscalar in determining the chirality, just like the parity violation in weak interactions~\cite{lee1956question,wu1957experimental}. The pseudoscalar is given by the dot product of a polar vector, related to some bond axis $\hat{\mathbf{n}}$, and an axial vector, reflecting the angle $\varphi$ between other two bonds with respect to $\hat{\mathbf{n}}$ in Fig.~\ref{chiral discrimination}(a). The spin angular momenta carried by the nearby atomic nuclei are axial vectors, and thus, it is quite natural to distinguish the chirality by relating nuclear spins to the angle $\varphi$ using the MR methods. The molecular system employed in this work is displayed in Fig.~\ref{chiral discrimination}(b) for illustrating how to determine the chirality-related angle through the spin-spin interactions. The system consists of the nitrogen atom, the vacancy and the nearby carbon atoms in diamond. For simplicity, only two $^{\text{13}}$C atoms carrying nuclear spins ($I=1/2$) are retained in Fig.~\ref{chiral discrimination}(b) given that $^{\text{12}}$C atoms are magnetic-silent. Two $^{\text{13}}$C spins have magnetic dipolar interactions with the electron spin ($S=1$) carried by the chiral center (the vacancy), with the difference of the azimuth angles $\Phi^{(12)} = \varphi^{(2)}-\varphi^{(1)}$ indicating the chirality. However, it is rather difficult to acquire the sign of $\Phi^{(12)}$ by means of the MR spectroscopy, in that the energy spectra for two enantiomers are identical, as shown in Fig.~\ref{chiral discrimination}(c). Only the eigenstates of nuclear spins in the subspace $\ket{m_S=+1}$ of the electron spin are rendered divergent by the sign of $\Phi^{(12)}$. To manifest the difference, the nuclear spins must experience different dynamics in both subspaces $\ket{m_S=+1}$ and $\ket{m_S=0}$ at the same time under the dynamical decoupling~\cite{de2010universal,taminiau2012detection} (DD) sequences. However, only the polar angle of the dipolar interaction takes effect in the DD results~\cite{taminiau2012detection,xie202399}, giving no information about the azimuth angle $\varphi^{(i)}$. To this end, the two-body dynamics of nuclear spins is induced by generating the indirect coupling~\cite{dutt2007quantum,nizovtsev2024indirect} between two nuclear spins via the electron spin in Fig.~\ref{chiral discrimination}(d), just like the J-coupling in conventional MR spectroscopy~\cite{bovey1988nuclear}. The interplay between single-body and two-body dynamics will preserve the azimuth angles into the final dynamical results in the form of an overall phase, $\Phi^{(12)} = \varphi^{(12)}-\varphi^{(1)}+\varphi^{(2)} \approx \varphi^{(2)}-\varphi^{(1)}$, where $\varphi^{(12)}$ is the phase of the flip-flop term of two nuclear spins (see Supplemental Material~\cite{sm}\nocite{xie2021identity, childress2006coherent, van1990electric, kucsko2013nanometre, blochl1994projector, kresse1993ab, kresse1994ab, perdew2008restoring, heyd2003hybrid, takacs2024accurate, neeseORCAProgramSystem2012}). The flip-flop phase $\varphi^{(12)}$ comes from the antisymmetric part of the J-coupling tensor, and is too small to be ignored in this study. As a matter of fact, there are some theoretical proposals~\cite{garbacz2016chirality,king2017antisymmetric} of leveraging the antisymmetric coupling to distinguish the chirality.

\begin{figure}
\centering
\begin{overpic}[width=1.0\columnwidth]{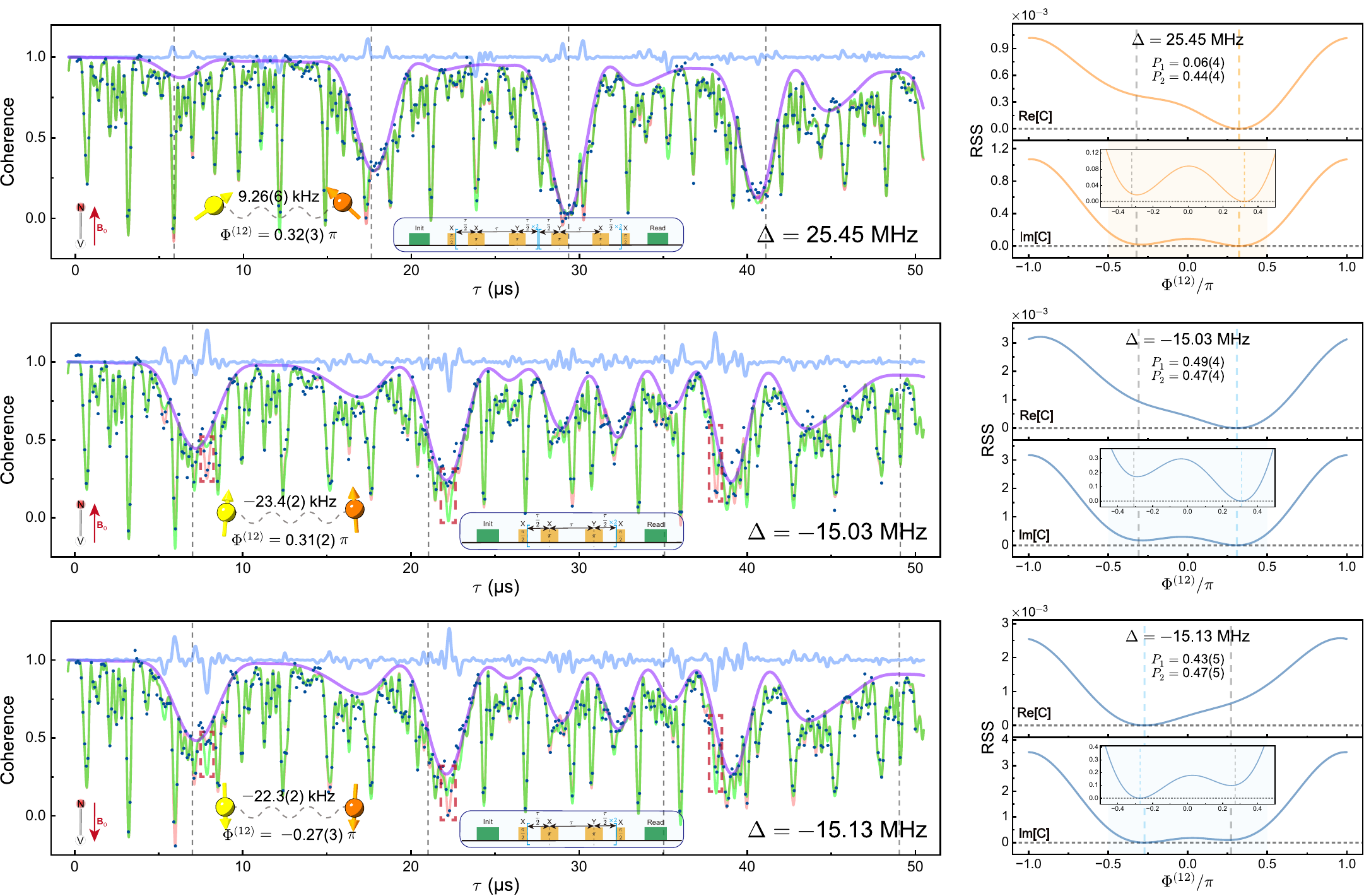}
	\put (-0.7, 64.1) {\figfont{(a)}}
	\put (-0.7, 42.1) {\figfont{(b)}}
	\put (-0.7, 20.3) {\figfont{(c)}}
	\put (70.2, 64.1) {\figfont{(d)}}
	\put (70.2, 42.1) {\figfont{(e)}}
	\put (70.2, 20.3) {\figfont{(f)}}
\end{overpic}
\caption{Chiral discrimination through symmetry-breaking dynamics of two nuclear spins under spin polarization. (a) Two-body dynamical results with the second spin partially polarized measured under the energy gap $\Delta$ = 25.45~MHz and the XY8 sequence. The red line is the fitting result with $\Phi^{(12)}$ = 0.32(3)~$\pi$ while the green one is plotted with $\Phi^{(12)}$ = $-$0.32~$\pi$. The blue one is the difference of these two lines with an offset of 1. (b) The dynamical results of two polarized spins obtained under $\Delta$ = $-$15.03~MHz and the XY4 sequence. (c) The dynamical results obtained by reversing the direction of the magnetic field. It is found that the results of two enantiomers are swapped by comparing the blue lines or the data in the red boxes in (b) and (c). (d-f) The calculated RSS based on the results under spin polarization in (a-c). $P_i$ is the polarization of the $i$th nuclear spin, and the field-inversion asymmetry is evidently manifested in (e) and (f).}\label{two-body}
\end{figure}

The experiments are performed near the ground-state level anticrossing~\cite{broadway2016anticrossing,wood2017microwave,broadway2018quantum} (GSLAC) with a small energy gap $\Delta$ between $\ket{m_S=-1}$ and $\ket{m_S=0}$, as shown in Fig.~\ref{chiral discrimination}(d). The NV center also acts as an MR detector for initializing and reading out the electron spin with laser illumination. First, the hyperfine coupling parameters of two nuclear spins are measured first by using the method of previous works~\cite{taminiau2012detection,xie202399} (see Supplemental Material~\cite{sm}). Then, a dynamical profile containing both sharp and wide dips in Fig.~\ref{chiral discrimination}(e) is measured under the energy gap $\Delta$ = 99.81~MHz by applying the XY8 sequence. The sharp dips mark the signature of single-body dynamics, while the emergence of wide dips~\cite{zhao2011atomic,shi2014sensing,abobeih2018one} results from the two-body dynamics driven by the flip-flop term of the J-coupling in the subspace $\ket{m_S=0}$ (Fig.~\ref{chiral discrimination}(c)). Fitting the entire results in Fig.~\ref{chiral discrimination}(e) outputs the flip-flop frequency $X^{(12)}$ = 2.49(3)~kHz and the chirality-related phase $\Phi^{(12)}$ = $\pm$0.33(4)~$\pi$ (see Supplemental Material~\cite{sm}). The dual-value result of the phase $\Phi^{(12)}$ means that two enantiomers experience the same dynamical process and are still indistinguishable. To break the mirror symmetry, the nuclear spins must be polarized, just like the parity-violation experiment~\cite{wu1957experimental}. Under $\Delta$ = 25.45~MHz, the second nuclear spin is evidently polarized due to the cross-relaxation process~\cite{jacques2009dynamic}, of which the polarization $P_1$ = 0.06(4) and $P_2$ = 0.44(4) is obtained by using the Ramsey sequence (see Supplemental Material~\cite{sm}). In this case, the symmetry has been already broken in the dynamical results of Fig.~\ref{two-body}(a), given that the unique value of the phase $\Phi^{(12)}$ = 0.32(3)~$\pi$ is obtained from the fitting and two curves with $\pm$0.32~$\pi$ are markedly different. Further narrowing the energy gap to $-$15.03~MHz gives rise to the polarization of both spins (0.49(4) and 0.47(4)), and a more evident divergence in the spin dynamics for two enantiomers in Fig.~\ref{two-body}(b). More spin dynamical results are recorded experimentally in the Supplemental Material~\cite{sm}. Furthermore, the absolute configuration is determined by measuring the direction $\hat{\mathbf{n}}^{\text{NV}}$ of the NV axis under a static electric field (see Supplemental Material~\cite{sm}). Besides, a magnetic-related enantiospecific effect is inevitably accompanied by the magnetic-field-inversion asymmetry~\cite{barron1986symmetry,luo2021chiral}. Such a prediction is experimentally validated in Fig.~\ref{two-body}(c) by reversing the magnetic field. The phase $\Phi^{(12)}$ is estimated to be $-$0.27(3)~$\pi$ with the sign changed, which means that the magnetic field reversal exchanges the spin dynamics for two enantiomers.

\begin{figure}
\centering
\centering
\begin{overpic}[width=1.0\columnwidth]{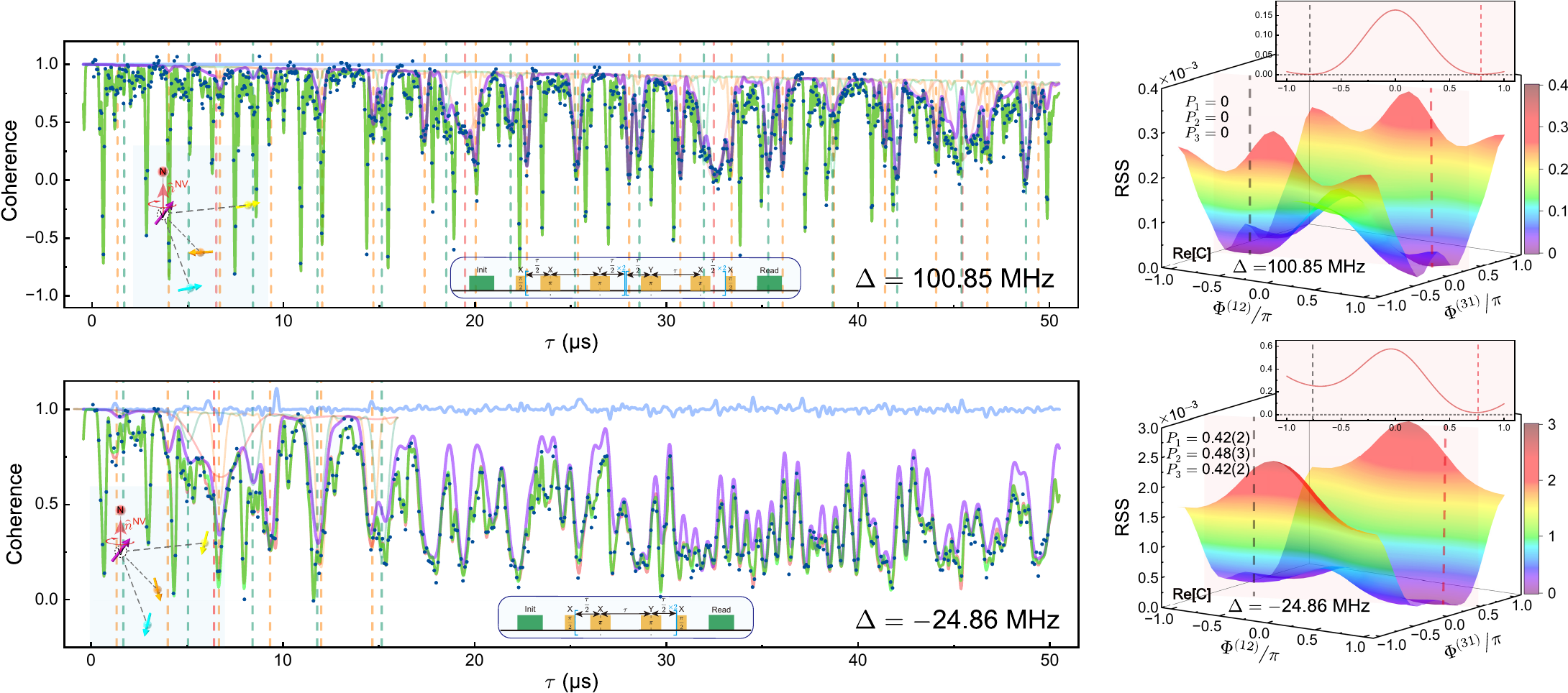}
	\put (-0.7, 41.9) {\figfont{(a)}}
	\put (-0.7, 20.1) {\figfont{(b)}}
	\put (70.3, 41.9) {\figfont{(c)}}
	\put (70.3, 20.1) {\figfont{(d)}}
\end{overpic}
\caption{Quantum three-body dynamics for chiral discrimination. (a) Three-body dynamics of unpolarized nuclear spins recorded under the energy gap $\Delta$ = 100.85~MHz by implementing the XY8 sequence. (b) The dynamical results of partially polarized nuclear spins obtained under $\Delta$ = $-$24.86~MHz and the XY4 sequence. The results of three pairs of two-body dynamics are calculated and plotted for the whole duration in (a) and the first 15~\textmu s in (b). (c,d) The RSS obtained from the three-body dynamical results in (a,b) in terms of $\Phi^{(12)}$ and $\Phi^{(31)}$. The inset plots a cross-section profile that contains the configurations of two enantiomers.}\label{three-body}
\end{figure}

Chiral discrimination is also demonstrated through probing and investigating quantum three-body dynamics of three nuclear spins. Under the energy gap $\Delta$ = 100.85~MHz, a dynamical profile for three unpolarized spins is recorded in Fig.~\ref{three-body}(a), and embraces more intricate structures than the two-body case in Fig.~\ref{two-body}. The dynamical behaviors in Fig.~\ref{three-body}(a) can be well understood by dividing three-body dynamics into three pairs of two-body dynamics for a short interval $\tau$. The whole results can be fitted nearly perfectly to give three flip-flop parameters ($X^{(12)}$ = 3.60(3)~kHz, $X^{(23)}$ = $-$2.94(4)~kHz and $X^{(31)}$ = $-$4.25(3)~kHz) and three phases ($\Phi^{(12)}$ = $\pm$0.79(10)~$\pi$, $\Phi^{(23)}$ = $\pm$0.95(13)~$\pi$ and $\Phi^{(31)}$ = $\pm$0.23(4)~$\pi$). Under a smaller gap $\Delta$ = $-$24.86~MHz, three nuclear spins are partially polarized (0.42(2), 0.48(3) and 0.42(2)), and the resultant dynamics is measured in Fig.~\ref{three-body}(b). The dynamical results appear more complicated compared with Fig.~\ref{three-body}(a), because three pairs of two-body dynamics gradually transition into a truly three-body dynamics. Indeed, spin polarization cancels the enantiomeric ambiguity by outputting the unique result for the phases (0.76(6)~$\pi$, 0.95(7)~$\pi$ and 0.32(3)~$\pi$) as well as the couplings ($-$17.24(11)~kHz, 14.46(12)~kHz and 21.03(11)~kHz).

An investigation of the symmetry-breaking mechanism is conducted here by performing the symmetry transformations on the nuclear spins. Under DD sequences, two nuclear spins with the initial state $\rho_n(P_1,P_2)$, experience different dynamical trajectories $U_0(\Phi^{(12)},A^{(12)}_{zz})$ and $U_1^{\dag}(\Phi^{(12)},A^{(12)}_{zz})$ under the action of the chirality-related angle $\Phi^{(12)}$ and the zz-component of the coupling tensor $A^{(12)}_{zz}$. The resultant coherence of the electron spin is given by,
\begin{align}
C(\Phi^{(12)},A^{(12)}_{zz},P_1,P_2) & = \text{Tr}\big[U_0(\Phi^{(12)},A_{zz}) \rho_n(P_1,P_2) U_1^{\dag}(\Phi^{(12)},A^{(12)}_{zz})\big] \label{coherence}\\
& = \text{Tr}\big[U_0^{\dag}(-\Phi^{(12)},A^{(12)}_{zz}) \rho_n(P_1,P_2) U_1(-\Phi^{(12)},A^{(12)}_{zz})\big]^{\ast}, \label{MT-symmetry}
\end{align}
where `\text{Tr}' means to take the trace of the matrix. Eq.~(\ref{MT-symmetry}) is derived by applying both the mirror reflection (M) and the time reversal (T) with the joint operator $MT=\kappa$ ($\kappa$ denotes complex conjugation), where $M=\sigma_y^{(1)} \otimes \sigma_y^{(2)}$ and $T=\sigma_y^{(1)} \otimes \sigma_y^{(2)}\kappa$ (see more details in the Supplemental Material~\cite{sm}). It can be inferred that the mirror symmetry in real space is correlated to the MT-symmetry in spin space, which is broken under spin polarization. To visualize the process of symmetry breaking, the deviation of the dynamical results with any other $\Phi^{(12)}$ from the real ones is calculated under different polarization. When the nuclear spins are unpolarized, the mirror symmetry in real space holds based on Eq.~(\ref{MT-symmetry}), as displayed in Fig.~\ref{chiral discrimination}(f). Along with the increase of the polarization, the degree of the symmetry breaking gradually steps up from Fig.~\ref{chiral discrimination}(f) to Fig.~\ref{two-body}(d,e), with the field-inversion symmetry breaking also exhibited in Fig.~\ref{two-body}(e,f). As for the three-body case in Fig.~\ref{three-body}, the symmetry breaking is manifested in Fig.~\ref{three-body}(c,d) with respect to two independent angles $\Phi^{(12)}$ and $\Phi^{(31)}$ by considering that the sum of three angles is an integer multiple of $2\pi$.

\begin{figure}
\centering
\begin{overpic}[width=1.0\columnwidth]{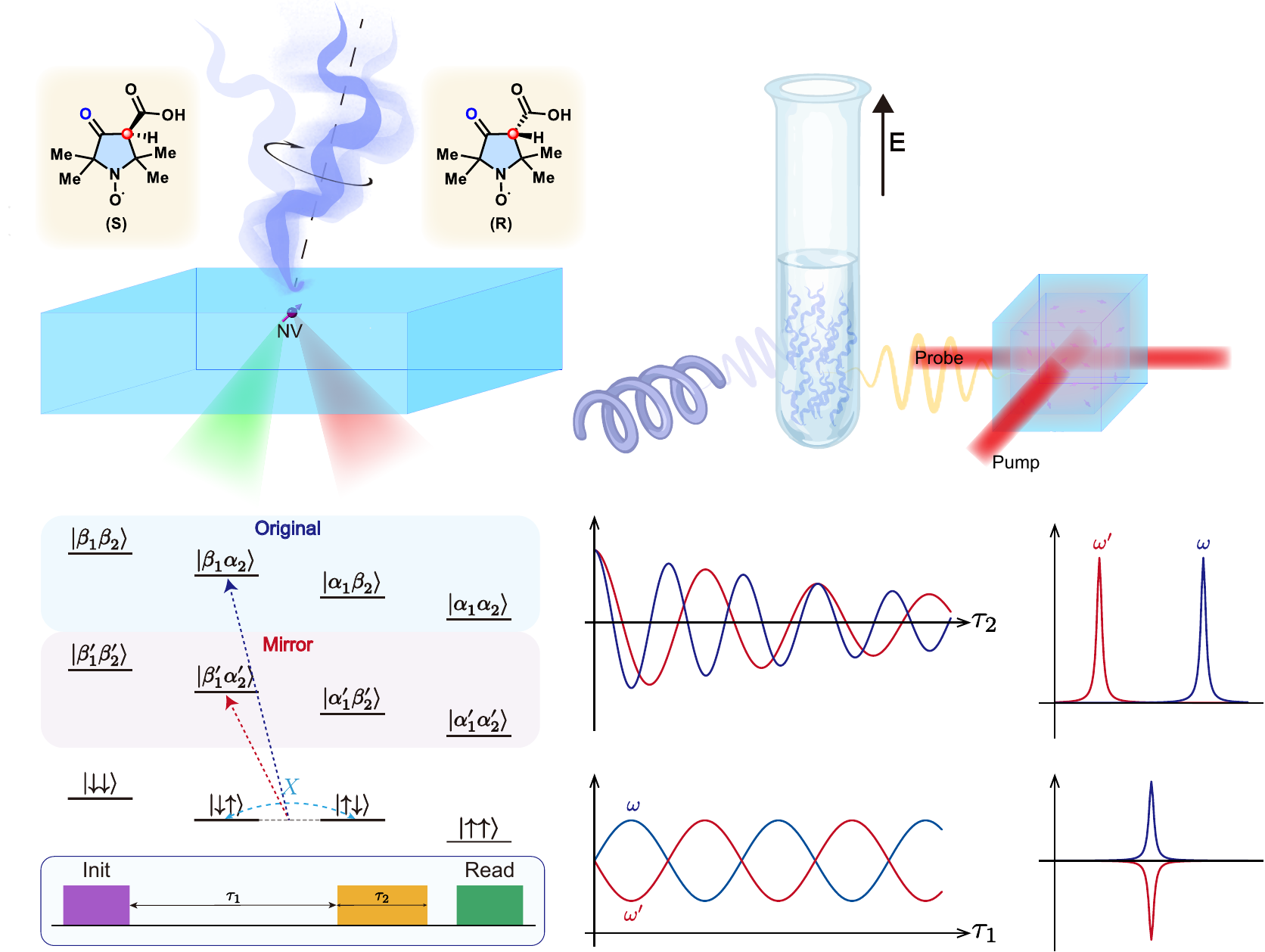}
	\put (0., 73.) {\figfont{(a)}}
	\put (48., 73.) {\figfont{(b)}}
	\put (0., 34.9) {\figfont{(c)}}
	\put (44., 34.9) {\figfont{(d)}}
\end{overpic}
\caption{Generalization and the underlying principle. (a) Setup for chiral discrimination of single chiral molecules with NV sensors. The molecule with an electron spin probe is swaying and rotating in solution above the diamond surface. Here a model molecule, 3-Carboxy-PROXYL, is listed as an example. (b) Extension of the method above to nuclear magnetic resonance under a low magnetic field for direct chiral discrimination of ensemble molecules. Molecular orientation is induced by an applied electric field to generate the residual dipolar coupling and the MR signal is detected by an optically pumped magnetometer. (c) Eigenstates in two subspaces for two enantiomers and the meta-process of the method for chiral discrimination. Here two sets of eigenstates are markedly different, as a special case, $\ket{\beta_1\alpha_2}=(\ket{\uparrow}+\ket{\downarrow})(\ket{\uparrow}+e^{i\Phi^{(12)}}\ket{\downarrow})$ and $\ket{\beta^\prime_1\alpha^\prime_2}=(\ket{\uparrow}+\ket{\downarrow})(\ket{\uparrow}+e^{-i\Phi^{(12)}}\ket{\downarrow})$. The system evolves under the flip-flop interaction for some time $\tau_1$, and then the transition strength to the $\ket{\beta_1\alpha_2}$ or $\ket{\beta^\prime_1\alpha^\prime_2}$ is measured by applying a selective pulse. (d) The results obtained under the sequence in (c). The transition strength is obtained by varying the time $\tau_2$, and exhibits opposite behaviors under the flip-flop dynamics for two enantiomers.}\label{generalization}
\end{figure}
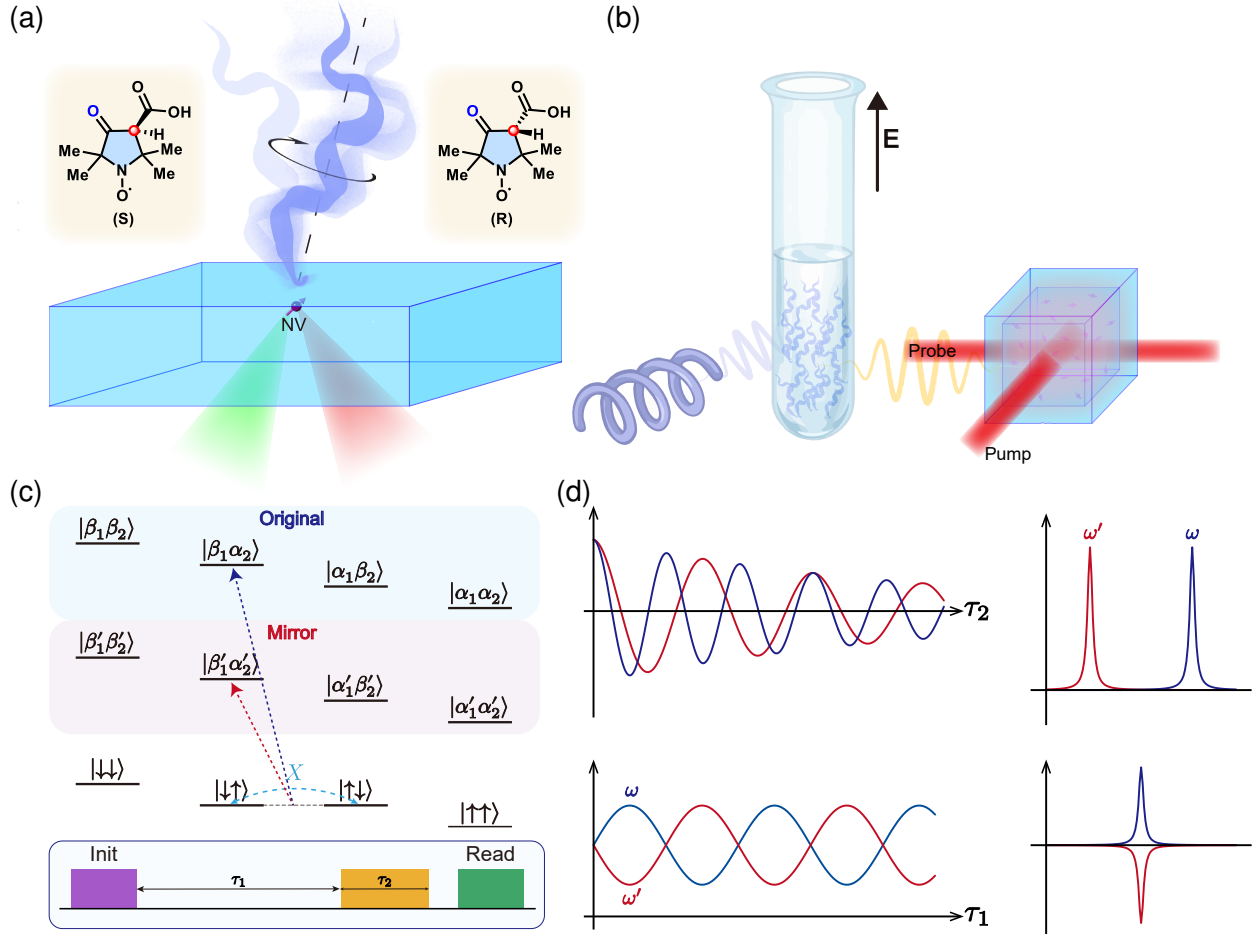

In the method above, the azimuth angles of dipolar interactions are tied together through the flip-flop dynamics of two spins to directly determine the chirality. Proposals in conventional nuclear magnetic resonance are focused on measuring the antisymmetric parts of the nuclear magnetic shielding polarizability~\cite{buckingham2004chirality,buckingham2006direct,walls2008measuring,buckingham2014communication,walls2014measuring} or the indirect spin-spin coupling tensor~\cite{walls2008measuring,garbacz2016chirality,king2017antisymmetric}, but these higher-order effects are quite small based on the theoretical calculations~\cite{garbacz2015theoretical,garbacz2016chirality}. In contrast, the dipolar interactions are ubiquitous and have a marked effect in magnetic resonance, which enables the method to be applied extensively. Figure~\ref{generalization}(a) portrays the scenario of directly detecting individual chiral molecules that are bonded to the diamond surface and rotating in solution. The method also works for the chiral discrimination of macroscopic samples under a low magnetic field (Fig.~\ref{generalization}(b)), where the residual dipolar coupling is induced by an electric field~\cite{riley2000extracting} and the magnetic signal is detected by optically pumped magnetometers~\cite{barskiy2025zero}. In fact, the dipolar interactions of the probed nuclear spins in real molecules might have roughly the same order of magnitude as the Zeeman splitting, since they are much closer to the central spin than those used here. As such, the eigenstates of two nuclear spins differ evidently in two subspaces, and a special case is displayed in Fig.~\ref{generalization}(c) for elucidating the underlying principle with $\ket{\beta_1\alpha_2}=(\ket{\uparrow}+\ket{\downarrow})(\ket{\uparrow}+e^{i\Phi^{(12)}}\ket{\downarrow})$ and $\Phi^{(12)}$ the chiral angle. In the lower subspace, two degenerate states evolve under the flip-flop interaction with $\cos{(X\tau_1/2)}\ket{\downarrow\uparrow}-i\sin{(X\tau_1/2)}\ket{\uparrow\downarrow}$, and then measuring its transition strength to the state $\ket{\beta_1\alpha_2}$, denoted by $1-\sin{(\Phi^{(12)})}\sin{(X\tau_1)}$, yields simple results in Fig.~\ref{generalization}(d) for distinguishing two enantiomers. In essence, it is the meta-process governing the dynamics under the DD sequence in Fig.~\ref{two-body} and Fig.~\ref{three-body}, where multiple $\pi$ pulses are needed to amplify the small difference of the eigenstates in two subspaces.

In conclusion, direct chiral determination is achieved experimentally at the single-molecule level by constructing the symmetry-breaking dynamics of the nuclear spins near the chiral center. The enantiomeric ambiguity is eliminated through probing the symmetry-breaking dynamics of nearby nuclear spins under spin polarization. The core of the method consists in correlating two azimuth angles of dipolar interactions through the flip-flop dynamics between two nuclear spins, leading to a measurable chirality-related angle $\Phi^{(12)}$. Since the method is not dependent on any weak higher-order effects but utilizes ordinary dipolar interactions, the method and the corresponding variation can be extensively applied in conventional magnetic resonance for direct chiral discrimination of macroscopic chiral samples. Furthermore, combined with the NV-based single-molecule MR method~\cite{du2024single}, the method opens up a broad space for revealing the underlying mechanisms of chirality-relevant phenomena at the single-molecule resolution in the future.

\section*{Acknowledgements}
This work was supported by the National Natural Science Foundation of China (grant nos. T2125011, 12274396, 12404555 and T2388102), the CAS (grant no. YSBR-068), the Innovation Program for Quantum Science and Technology (grants nos. 2021ZD0302200 and 2021ZD0303204), the China Postdoctoral Science Foundation (grant nos. 2023M743399 and GZB20240718), and the Fundamental Research Funds for the Central Universities, New Cornerstone Science Foundation through the XPLORER PRIZE.

\clearpage
\bibliography{references.bib}

\clearpage

\clearpage

\clearpage

\clearpage

\end{document}


\title{Supplemental Material for \\ Molecular chiral discrimination through symmetry-breaking spin dynamics}

\author{Tianyu Xie}
\thanks{T. Xie and Y. Hao contributed equally to this work.}
\affiliation{Laboratory of Spin Magnetic Resonance, School of Physical Sciences, Anhui Province Key Laboratory of Scientific Instrument Development and Application, University of Science and Technology of China, Hefei 230026, China}
\affiliation{Hefei National Research Center for Physical Sciences at the Microscale, University of Science and Technology of China, Hefei 230026, China}

\author{Yucheng Hao}
\thanks{T. Xie and Y. Hao contributed equally to this work.}
\affiliation{Laboratory of Spin Magnetic Resonance, School of Physical Sciences, Anhui Province Key Laboratory of Scientific Instrument Development and Application, University of Science and Technology of China, Hefei 230026, China}
\affiliation{Hefei National Research Center for Physical Sciences at the Microscale, University of Science and Technology of China, Hefei 230026, China}

\author{Mingzhe Liu}
\affiliation{Laboratory of Spin Magnetic Resonance, School of Physical Sciences, Anhui Province Key Laboratory of Scientific Instrument Development and Application, University of Science and Technology of China, Hefei 230026, China}

\author{Mengqi Wang}
\affiliation{Laboratory of Spin Magnetic Resonance, School of Physical Sciences, Anhui Province Key Laboratory of Scientific Instrument Development and Application, University of Science and Technology of China, Hefei 230026, China}
\affiliation{Hefei National Research Center for Physical Sciences at the Microscale, University of Science and Technology of China, Hefei 230026, China}

\author{Shaoyi Xu}
\affiliation{Laboratory of Spin Magnetic Resonance, School of Physical Sciences, Anhui Province Key Laboratory of Scientific Instrument Development and Application, University of Science and Technology of China, Hefei 230026, China}

\author{Zhiyuan Zhao}
\affiliation{Laboratory of Spin Magnetic Resonance, School of Physical Sciences, Anhui Province Key Laboratory of Scientific Instrument Development and Application, University of Science and Technology of China, Hefei 230026, China}
\affiliation{Hefei National Research Center for Physical Sciences at the Microscale, University of Science and Technology of China, Hefei 230026, China}

\author{Chang-Kui Duan}
\affiliation{Laboratory of Spin Magnetic Resonance, School of Physical Sciences, Anhui Province Key Laboratory of Scientific Instrument Development and Application, University of Science and Technology of China, Hefei 230026, China}
\affiliation{Hefei National Research Center for Physical Sciences at the Microscale, University of Science and Technology of China, Hefei 230026, China}
\affiliation{Hefei National Laboratory, University of Science and Technology of China, Hefei 230088, China}

\author{Ya Wang}
\affiliation{Laboratory of Spin Magnetic Resonance, School of Physical Sciences, Anhui Province Key Laboratory of Scientific Instrument Development and Application, University of Science and Technology of China, Hefei 230026, China}
\affiliation{Hefei National Research Center for Physical Sciences at the Microscale, University of Science and Technology of China, Hefei 230026, China}
\affiliation{Hefei National Laboratory, University of Science and Technology of China, Hefei 230088, China}

\author{Fazhan Shi}
\email{fzshi@ustc.edu.cn}
\affiliation{Laboratory of Spin Magnetic Resonance, School of Physical Sciences, Anhui Province Key Laboratory of Scientific Instrument Development and Application, University of Science and Technology of China, Hefei 230026, China}
\affiliation{Hefei National Research Center for Physical Sciences at the Microscale, University of Science and Technology of China, Hefei 230026, China}
\affiliation{Hefei National Laboratory, University of Science and Technology of China, Hefei 230088, China}
\affiliation{School of Biomedical Engineering and Suzhou Institute for Advanced Research, University of Science and Technology of China, Suzhou 215123, China}
\affiliation{The First Affiliated Hospital of USTC, Division of Life Sciences and Medicine, University of Science and Technology of China, Hefei 230026, China}

\author{Jiangfeng Du}
\affiliation{State Key Laboratory of Ocean Sensing and School of Physics, Zhejiang University, Hangzhou 310058, China}

\maketitle

\section*{Diamond sample}
The diamond sample used in the experiments is synthesized through chemical vapor deposition, whose top face is perpendicular to the [100] crystal axis and lateral faces are perpendicular to [110]. The nitrogen concentration of the diamond is less than 5 parts per billion and the abundance of $^{\text{13}}$C atoms is at the natural level of 1.1\%. Individual NV centers are created by 10-MeV electron irradiation to a total dose of 1.0~kGy and subsequent annealing at 1000~\degree C for 4 hours in vacuum. The negatively charged NV (NV$^-$) centers with the ground-state spin triplet are employed in this work for generating the indirect coupling for the surrounding $^{\text{13}}$C nuclear spins.

\section*{Experimental setup}
\begin{figure}
\centering
\includegraphics[width=1.0\textwidth]{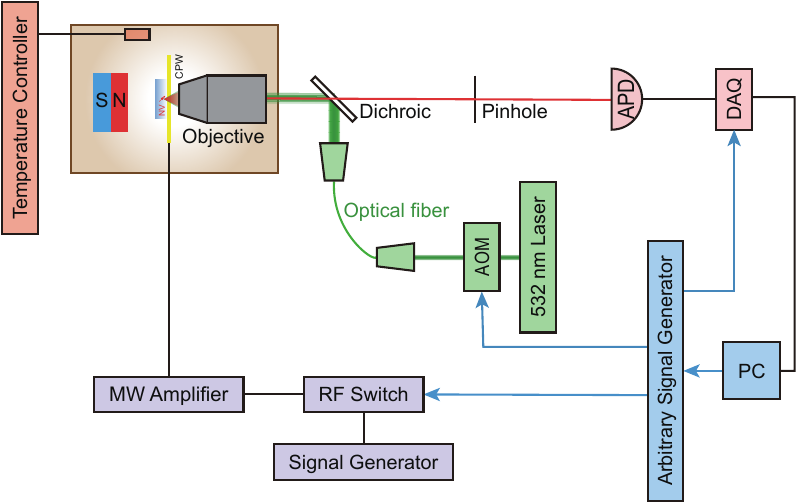}
\caption{Experimental setup for optically detected magnetic resonance. The setup is mainly comprised of microwave circuit (purple), optical system for excitation (green) and imaging (red), synchronization system (blue), and a diamond platform
inside a temperature-controlled box.}\label{setup}
\end{figure}

The schematic illustration of our experimental setup is shown in Fig.~\ref{setup}. All the experiments are performed on a home-built optically detected magnetic resonance platform. The instructions created by the computer are loaded into the arbitrary signal generator (CIQTEK, ASG8005) for controlling the whole system to work in an orderly manner. The 532-nm laser used for initializing and reading out the spin state of the NV center, is directed twice through acousto-optic modulators (Gooch \& Housego, contrast ratio $\sim$ 1/1000), and then fed into the wavelength division multiplexer (Thorlabs, RGB46HA). The fluorescence photons emitted from the NV center are collected by an oil objective (Olympus, UPLXAPO 100XO, NA 1.45), and detected by an avalanche photodiode (APD) (Perkin Elmer, SPCM-AQRH-14). The data acquisition card (National Instruments, 6363) installed on the computer counts the output signals from the APD. The microwave pulses are generated by a signal generator (Rigol ADG3065B) combined with a pin diode switch (Custom Microwave Components, CMCS0947A-C2), amplified by a high-power amplifier (Talent Microwave, THPA2G18G-43-43BC-Gain), and finally delivered into a coplanar waveguide (CPW) for manipulating the NV spin state. The diamond sample is mounted on the CPW and positioned by a piezo flexure scanner, while a permanent magnet is mounted on a three-dimensional positioner to provide a bias field along the NV axis. The magnet and positioner, the objective, and the piezo flexure scanner together with the diamond and CPW, are all placed inside a thermal insulation copper box. The temperature inside the box is stabilized at $\sim$~mK level through the PID feedback of the temperature controller (Stanford, PTC10).

\section*{Magnetic field alignment and stability}
\begin{figure}
\centering
\begin{overpic}[width=1.0\textwidth]{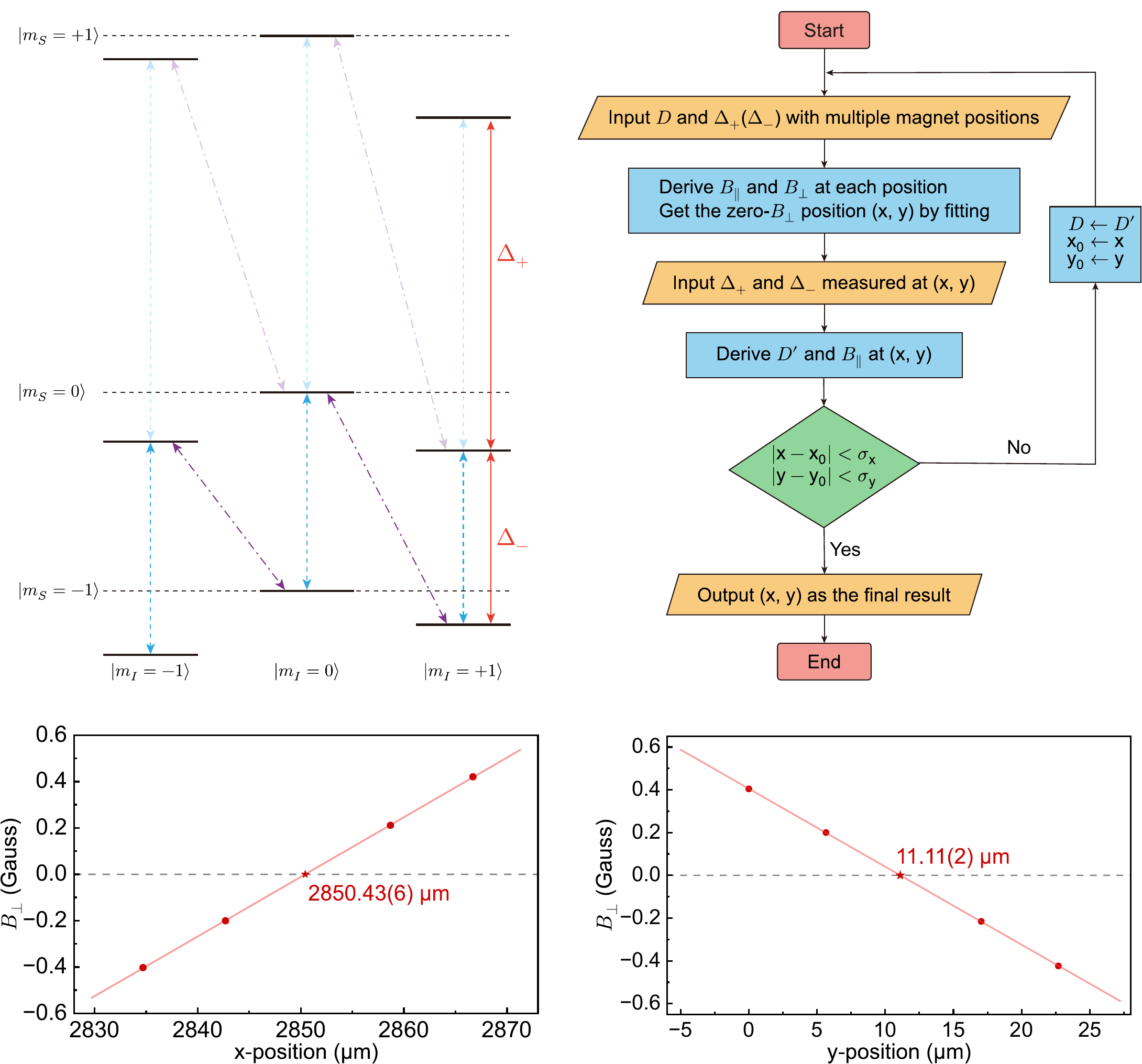}
	\put (-0.7, 91.5) {\figfont{(a)}}
	\put (50., 91.5) {\figfont{(b)}}
	\put (-0.7, 28.9) {\figfont{(c)}}
	\put (51.5, 28.9) {\figfont{(d)}}
\end{overpic}
\caption{Magnetic field alignment. (a) Level diagram of the coupled electron and $^{\text{14}}$N nuclear spins. The dashed-dotted (purple) and dashed (blue) lines describe the matrix elements from the transverse term $A^{\text{N}}_{\perp}$ of the $^{\text{14}}$N hyperfine tensor and the Zeeman effect caused by the transverse field $B_\perp$. The solid lines (red) in the subspace $\ket{m_I=+1}$ mark two energy gaps $\Delta_+$ and $\Delta_-$ that are measured experimentally. (b) Flow diagram of the iterative algorithm to perform magnetic field alignment. The values of $B_\parallel$ and $B_\perp$ are derived by solving the whole Hamiltonian of the coupled spin system. (c,d) The derived $B_\perp$ as a function of the x and y position of the magnet after the value of $D$ is optimized. The zero-$B_\perp$ position is determined by a linear fit.}\label{alignment}
\end{figure}

Aligning the magnetic field along the NV axis is realized by fitting the transverse field $B_{\perp}$ with the magnet position to find the zero-$B_{\perp}$ position. The transverse field $B_{\perp}$ at each position is obtained by measuring two resonant frequencies ($\Delta_+$ and $\Delta_-$) and solving the nine-level Hamiltonian of the $^{\text{14}}$N$-$NV system~\cite{xie2021identity}, as shown in Fig.~\ref{alignment}(a). The entire process is detailed in Fig.~\ref{alignment}(b): 1. An initial value for the zero-field splitting $D$ of the electron spin is obtained by measuring $\Delta_+$ and $\Delta_-$ far away from the GSLAC; 2. Move to a certain position near the GSLAC, and measure $\Delta_+$ and $\Delta_-$. Two components of the magnetic field $B_{\parallel}$ and $B_{\perp}$ can be obtained by solving the nine-level Hamiltonian with the values of $D$, $\Delta_+$ and $\Delta_-$. Acquire multiple sets of $B_{\parallel}$ and $B_{\perp}$ at different magnet positions, and then the zero-$B_{\perp}$ position (x, y) can be pinpointed by fitting $B_{\perp}$ with the magnet position; 3. Move to (x, y), and measure $\Delta_+$ and $\Delta_-$. The value of $D$ is updated by solving the nine-level Hamiltonian with $B_{\perp}=0$; 4. Repeat the steps above until the variation of the zero-$B_{\perp}$ position is smaller than the positioning accuracy (300~nm) of the magnet positioner. As an example, the results for determining the zero-$B_{\perp}$ position when $\Delta_-$ $\approx$ 5.7~MHz are exhibited in Fig.~\ref{alignment}(c) and Fig.~\ref{alignment}(d). The $B_{\perp}$ gradients with respect to the x and y directions are measured to be 0.02573(13)~G/\textmu m and $-$0.03646(10)~G/\textmu m by linearly fitting the results. The residual $B_{\perp}$ is estimated to be $\sim$ 0.01~G by considering the positioning accuracy. Besides, the thermal insulation box in Fig.~\ref{setup} is placed within an acrylic sheet cabinet attached by heat-insulating foam to improve the field stability. The temperature stability is controlled within $\pm0.5$~K in the lab by an air conditioner, $\pm0.2$~K in the cabinet, and $\pm1$~mK in the thermal insulation box. As a result, the long-term stability of the magnetic field is maintained within $\pm0.04$~G, corresponding to $\pm100$~kHz for $\Delta_-$.

\section*{Theoretical derivations for the indirect coupling between two nuclear spins}
At first, we derive the indirect coupling between two nuclear spins via the method of canonical transformation, which is often used in condensed matter physics. As a starting point, the whole Hamiltonian can be partitioned into two parts,
\begin{equation}
H = H_0 + H_1,
\label{Hamiltonian}
\end{equation}
where $H_0$ is the principal term for determining the energy levels of the system, and $H_1$ is the perturbation term containing the matrix elements between different energy levels. To remove the perturbation term $H_1$, a unitary transformation with a small anti-Hermitian operator $S$,
\begin{equation}
U = e^S,
\label{unitary}
\end{equation}
is performed onto the Hamiltonian $H$, leading to
\begin{equation}
H^\prime = e^S H e^{-S} = H_0 + H_1 + [S,H_0] + [S,H_1] + \frac{1}{2!}[S,[S,H_0]] + \cdots, 
\label{Hamiltonian_2}
\end{equation}
where $[,]$ denotes the commutator with $[A,B] = AB-BA$. To eliminate $H_1$ in the new picture, the equation
\begin{equation}
H_1 + [S, H_0] = 0
\label{operator_equation}
\end{equation}
is forced to hold, from which the operator $S$ is determined. As such, an approximate expression of the Hamiltonian $H^\prime$ can be denoted as
\begin{equation}
H^\prime \approx H_0 + \frac{1}{2}[S,H_1],
\label{Hamiltonian_f}
\end{equation}
by truncating it to the second-order term. A new term $\frac{1}{2}[S, H_1]$ appears after removing $H_1$, and may contain an indirect interaction between two particles.

By applying the theoretical framework above to the hybrid spin system in this work, where two nuclear spins have hyperfine interactions with an electron spin, an indirect interaction between two nuclear spins will be induced with dependence on the energy gap. The principal term of the whole Hamiltonian with an external field applied along the NV axis is written as
\begin{equation}
H_0 = D S_z^2 + \omega_e S_z + \omega_{\text{L}} (I_z^{(1)} + I_z^{(2)}),
\label{Hamiltonian_0}
\end{equation}
where $S_z$, $I_z^{(1)}$, and $I_z^{(2)}$ are the third components of the spin operators ($\mathbf{S}$, $\mathbf{I}^{(1)}$, and $\mathbf{I}^{(2)}$) of the electron spin, the first nuclear spin, and the second nuclear spin, respectively. $D$ is the zero-field splitting of the NV electron spin, $\omega_e$ is the Zeeman splitting of the electron spin, and $\omega_{\text{L}}$ is the Larmor precession frequency of $^{\text{13}}$C nuclear spins $\approx$ $-$1.1~MHz near the GSLAC. The perturbation term $H_1$ comes from the hyperfine interactions of two nuclear spins with the electron spin,
\begin{equation}
H_{\text{hp}} = \sum_{i=1}^2 \mathbf{S}\cdot\mathbf{A}^{(i)}\cdot\mathbf{I}^{(i)} = \sum_{i=1}^2 S_z\mathbf{A}^{(i)}_z\cdot\mathbf{I}^{(i)} + S_x\mathbf{A}^{(i)}_x\cdot\mathbf{I}^{(i)} +
S_y\mathbf{A}^{(i)}_y\cdot\mathbf{I}^{(i)},
\label{hyperfine}
\end{equation}
where the hyperfine tensors can be explicitly represented as
\begin{equation}
\mathbf{A}^{(i)} =
\begin{pmatrix}
    \mathbf{A}^{(i)}_{x} \\
    \mathbf{A}^{(i)}_{y} \\
    \mathbf{A}^{(i)}_{z} \\
\end{pmatrix} =
\begin{pmatrix}
    A^{(i)}_{xx} & A^{(i)}_{xy} & A^{(i)}_{xz} \\
    A^{(i)}_{yx} & A^{(i)}_{yy} & A^{(i)}_{yz} \\
    A^{(i)}_{zx} & A^{(i)}_{zy} & A^{(i)}_{zz} \\
\end{pmatrix}.
\end{equation}
The terms in the third row ($\mathbf{A}_z^{(i)}$), defined as secular terms, commute with $H_0$, and thus can be incorporated into $H_0$. Actually, the perturbation term $H_1$ is composed of the first two rows of the hyperfine tensors ($\mathbf{A}_x^{(i)}$ and $\mathbf{A}_y^{(i)}$), defined as non-secular terms.

Near the GSLAC, two energy levels $\ket{m_S=0}$ and $\ket{m_S=-1}$ are close, where the indirect coupling of nuclear spins is generated, and the third level $\ket{m_S=+1}$ is about 5.7~GHz far away and ignored in the following derivations. Compared with the energy gap $\Delta$ between $\ket{m_S=0}$ and $\ket{m_S=-1}$, the energy levels of nuclear spins in every subspace are considered to be degenerate, so that the operator $S$ can be determined by Eq.~(\ref{operator_equation}),
\begin{equation}
P_{m_S=0} S P_{m_S=-1} = \frac{1}{\Delta} P_{m_S=0} H_1 P_{m_S=-1},
\label{S_operator}
\end{equation}
where $P_{m_S=0}$ and $P_{m_S=-1}$ are the projection operators of two subspaces $\ket{m_S=0}$ and $\ket{m_S=-1}$. The induced Hamiltonian in the subspace $\ket{m_S=0}$ after the canonical transformation is given by
\begin{equation}
\begin{split}
H^{\prime}_{m_S=0} & = \frac{1}{2} P_{m_S=0} [S,H_1] P_{m_S=0} = \frac{1}{\Delta} P_{m_S=0} H_1 P_{m_S=-1} H_1 P_{m_S=0}, \\
& = \sum_{i=1}^2 \frac{\lvert\mathbf{A}_x^{(i)}\rvert^2 + \lvert\mathbf{A}_y^{(i)}\rvert^2}{8\Delta} - \frac{\mathbf{A}_x^{(i)} \times \mathbf{A}_y^{(i)} \cdot \mathbf{I}^{(i)}}{2\Delta} + \mathbf{I}^{(1)}\cdot\mathbf{A}^{(12)}\cdot\mathbf{I}^{(2)},
\label{H0_induced}
\end{split}
\end{equation}
where the third term is the induced coupling between two nuclear spins with the coupling tensor~\cite{dutt2007quantum} given by
\begin{equation}
A^{(12)}_{ij} = \frac{1}{\Delta} \sum_{k=x,y} A^{(1)}_{ki} A^{(2)}_{kj}.
\label{natural}
\end{equation}
The first term is the energy shift of the whole subspace and ignored in this work. The second term introduces an extra effective magnetic field in the subspace $\ket{m_S=0}$, which plays an important role in the theoretical model below for describing the dynamics of nuclear spins. As a matter of fact, the flip-flop strength given by Eq.~(\ref{natural}) is somewhat inaccurate, because the levels of two $^{\text{13}}$C nuclear spins in every subspace are not degenerate enough and the electron spin states are mixed by the coupling with the $^{\text{14}}$N nuclear spin~\cite{xie2021identity}. Nevertheless, it does not matter since the expression of Eq.~(\ref{natural}) is not used in this work and the flip-flop strength is obtained by fitting the data.

\section*{Spin dynamics under the DD sequences}
In this work, the dynamical behaviors of multiple $^{\text{13}}$C nuclear spins, including single-body, two-body and three-body dynamics, are all recorded into the coherence of the electron spin under the DD sequences~\cite{childress2006coherent,taminiau2012detection}. Microwave pulses of the DD sequences are played between the state $\ket{m_S=0}$ and $\ket{m_S=+1}$ of the electron spin. To calculate the state evolution under the DD sequences, the entire Hamiltonian of the electron spin and nuclear spins needs to be first transformed as
\begin{equation}
H = \ket{0}\bra{0} \otimes H_{0} + \ket{+1}\bra{+1} \otimes H_{1},
\label{hamiltonian_DD}
\end{equation}
where $H_{0}$ ($H_{1}$) is the reduced Hamiltonian of nuclear spins in the subspace $\ket{m_S=0}$ ($\ket{m_S=+1}$). Then, it is not hard to calculate the state evolution under the Hamiltonian of Eq.~(\ref{hamiltonian_DD}),
\begin{equation}
U = \ket{0}\bra{0} \otimes U_{0} + \ket{+1}\bra{+1} \otimes U_{1}.
\label{unitary_DD}
\end{equation}
For an even number of $\pi$ pulses $N_\pi$, $U_0$ and $U_1$ are given by
\begin{equation}
U_{0} = (u_{0} u_{1}^2 u_{0})^{N_\pi/2},\text{ } U_{1} = (u_{1} u_{0}^2 u_{1})^{N_\pi/2},
\label{U0U1}
\end{equation}
and for the Hahn echo sequence with only one $\pi$ pulse,
\begin{equation}
U_{0} = u_{1} u_{0},\text{ } U_{1} = u_{0} u_{1},
\label{U0U1_echo}
\end{equation}
where $u_0$ ($u_1$) is the unit evolution of nuclear spins in the subspace $\ket{m_S=0}$ ($\ket{m_S=+1}$),
\begin{equation}
u_{0} = e^{-i H_0 \tau/2},\text{ } u_{1} = e^{-i H_{1} \tau/2},
\label{u0u1}
\end{equation}
An even superposition of $\ket{m_S=0}$ and $\ket{m_S=+1}$ of the electron spin is prepared by the first $\pi/2$ pulse, while the final $\pi/2$ pulse reads the real part of the coherence of the electron spin as
\begin{equation}
C_{\text{r}} = \text{Re}[\text{Tr}(U_{0} \rho_n U_{1}^{\dag})],
\label{coherence_r}
\end{equation}
where $\rho_n$ is the initial state of multiple nuclear spins after optical pumping. From Eq.~(\ref{U0U1}), it is obvious that if $H_0$ commutes with $H_1$, $U_0 = U_1$ and the coherence will be constantly one. Therefore, the information contained in the coherence requires the non-commutativity of $H_0$ and $H_1$, which is satisfied for all dynamical behaviors of nuclear spins in this work.

\section*{The hyperfine parameters of the NV centers used in the experiments}
\begin{figure}
\centering
\begin{overpic}[width=1.0\textwidth]{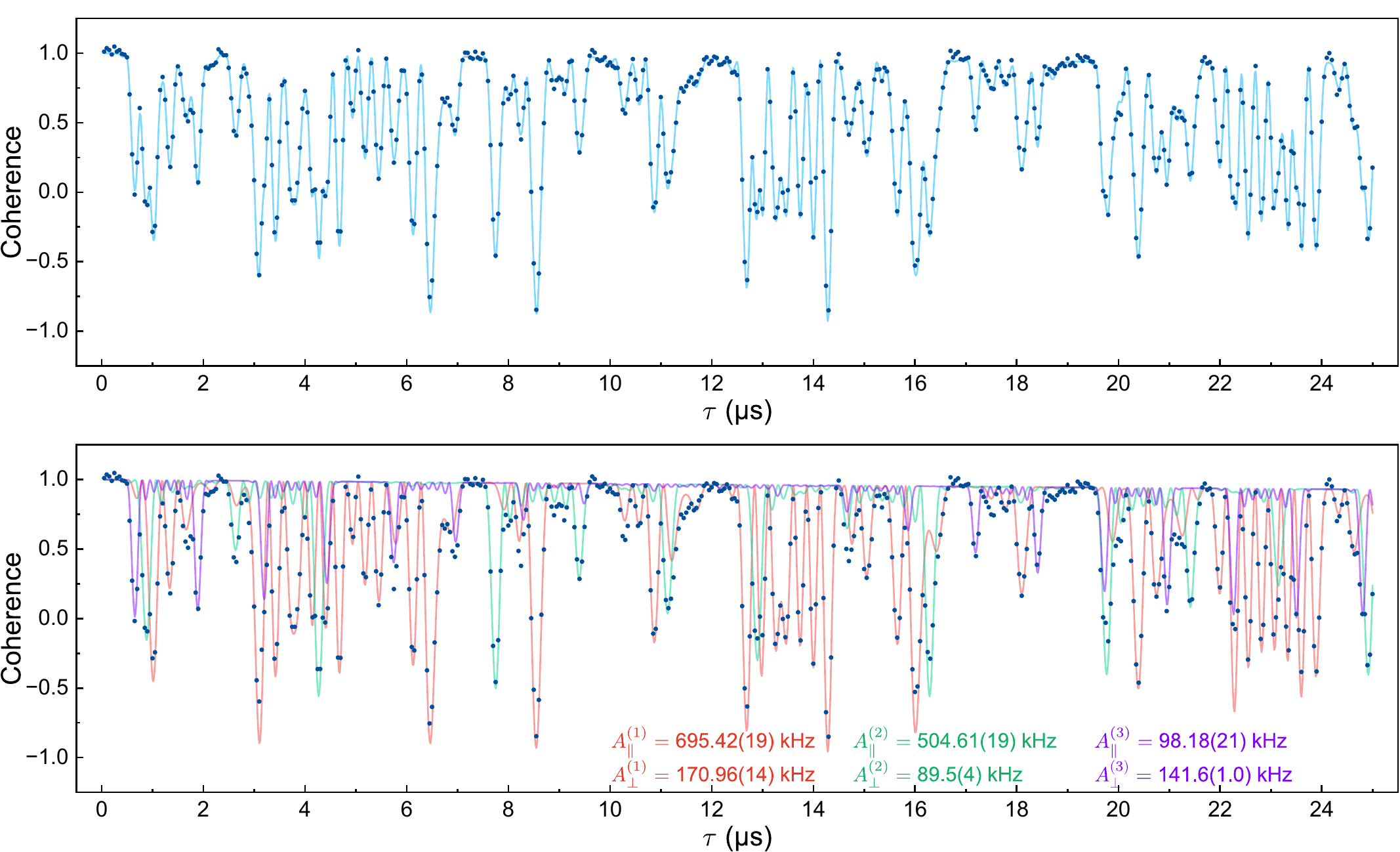}
	\put (-0.7, 59.7) {\figfont{(a)}}
	\put (-0.7, 29.1) {\figfont{(b)}}
\end{overpic}
\caption{Resolving $^{\text{13}}$C nuclear spins for chiral discrimination with two-body dynamics. (a) The decoherence profile of the electron spin under the energy gap $\Delta$ = $-$702.28~MHz by applying the XY8 sequence. The results are fitted by combining the decoherence contributions (blue line) from three nuclear spins and spin bath to give the hyperfine parameters in (b). (b) The decoherence of the electron spin caused by three nuclear spins. Three curves are plotted with the hyperfine parameters in the inset. The third nuclear spin does not take effect in the dynamical results in Fig.~2 of the main text.}\label{two-body_13C}
\end{figure}

\begin{figure}
\centering
\begin{overpic}[width=1.0\textwidth]{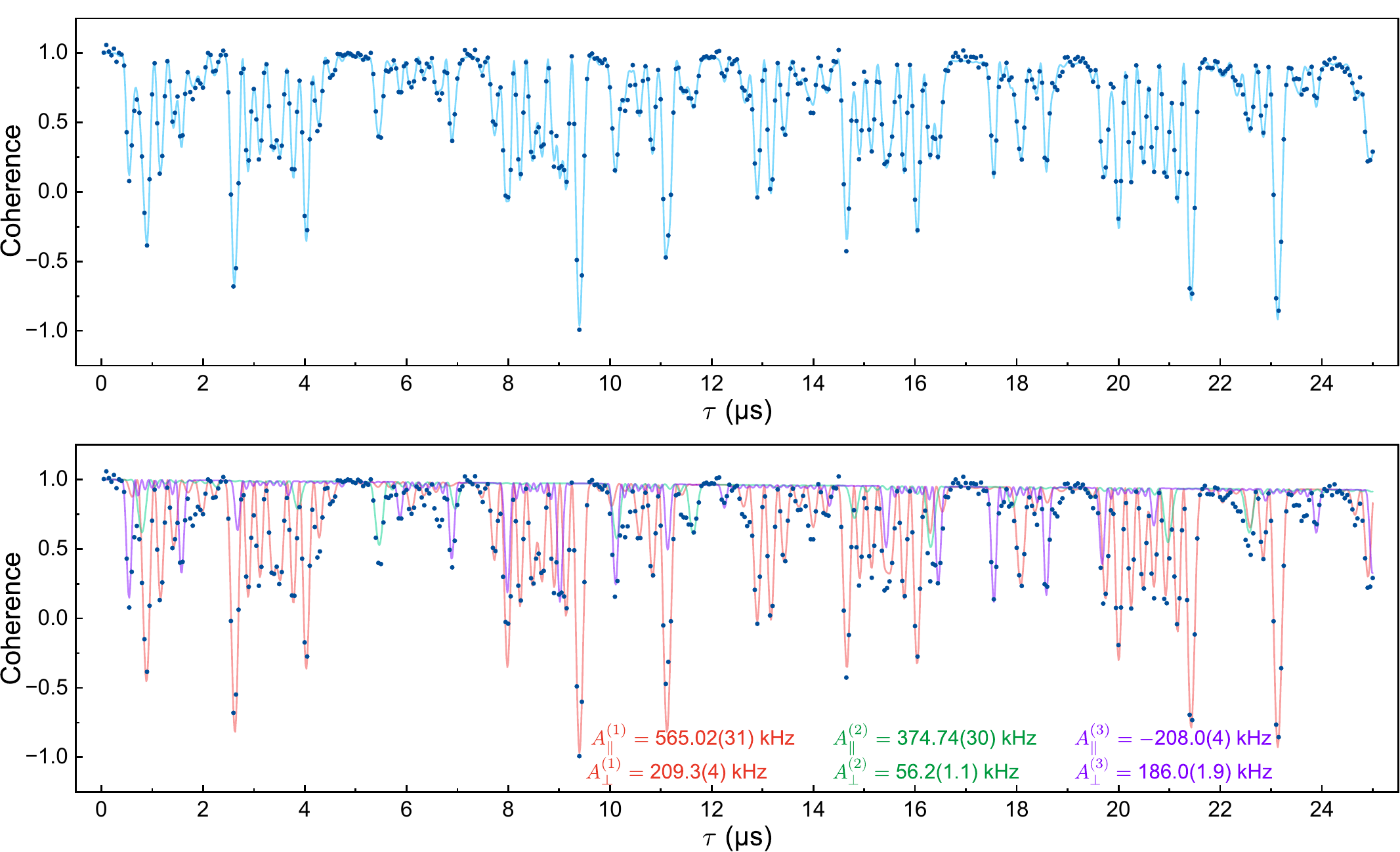}
	\put (-0.7, 59.7) {\figfont{(a)}}
	\put (-0.7, 29.1) {\figfont{(b)}}
\end{overpic}
\caption{Resolving $^{\text{13}}$C nuclear spins for chiral discrimination with three-body dynamics. (a) The decoherence profile of the electron spin under the energy gap $\Delta$ = $-$702.32~MHz by applying the XY8 sequence. The blue line is the total decoherence from three nuclear spins and spin bath with the hyperfine parameters given in (b). (b) The decoherence of the electron spin caused by three nuclear spins. Three curves are plotted with the hyperfine parameters in the inset.}\label{three-body_13C}
\end{figure}

The NV centers used for chiral discrimination through symmetry-breaking two-body and three-body dynamics in the main text are described in this section. The hyperfine parameters of the $^{\text{13}}$C nuclear spins need to be measured by adopting the method of previous works~\cite{taminiau2012detection,xie202399}. The experiments are performed under a large energy gap $\Delta$ $\approx$ $-$700~MHz by applying the DD sequences, where the J-couplings between nuclear spins are negligible. Therefore, it is just a simple application of Eq.~(\ref{coherence_r}) for the single-body dynamics of multiple nuclear spins with the Hamiltonian in the subspace $\ket{m_S=0}$ and $\ket{m_S=+1}$ given by
\begin{equation}
H_0 = \sum_{i=1}^{N} \omega_\text{L} I_z^{(i)},\text{ } H_1 = \sum_{i=1}^{N} (\omega_\text{L}+A_\parallel^{(i)}) I_z^{(i)} + A_\perp^{(i)}(\cos{\varphi^{(i)}}I_x^{(i)} + \sin{\varphi^{(i)}}I_y^{(i)}),
\label{H0H1_single_0}
\end{equation}
where $A_\parallel^{(i)}=A_{zz}^{(i)}$, $A_\perp^{(i)}\cos{\varphi^{(i)}}=A_{zx}^{(i)}$, and $A_\perp^{(i)}\sin{\varphi^{(i)}}=A_{zy}^{(i)}$. The azimuth angles $\varphi^{(i)}$ of nuclear spins are redundant for calculating the coherence of Eq.~(\ref{coherence_r}), and thus $H_0$ and $H_1$ can be simplified into
\begin{equation}
H_0 = \sum_{i=1}^{N} \omega_\text{L} I_z^{(i)},\text{ } H_1 = \sum_{i=1}^{N} (\omega_\text{L}+A_\parallel^{(i)}) I_z^{(i)} + A_\perp^{(i)} I_x^{(i)}.
\label{H0H1_single}
\end{equation}
Following the practice of this work~\cite{taminiau2012detection}, the Hamiltonian can be further represented with two non-parallel effective fields in two subspaces,
\begin{equation}
H_0 = \sum_{i=1}^{N} \omega_0^{(i)} I_z^{(i)},\text{ } H_1 = \sum_{i=1}^{N} \omega_1^{(i)} \mathbf{n}^{(i)} \cdot \mathbf{I}^{(i)},
\label{H0H1_single_f}
\end{equation}
where $\omega_0^{(i)}=\omega_\text{L}$, $\omega_1^{(i)}=-\sqrt{(\omega_\text{L}+A_\parallel^{(i)})^2+(A_\perp^{(i)})^2}$, and $\mathbf{n}^{(i)}=(A_\perp^{(i)},0,\omega_\text{L}+A_\parallel^{(i)})/\omega_1^{(i)}$. Nuclear spins are unpolarized under the energy gap $\Delta$ $\approx$ 700~MHz, so the initial state of nuclear spins are taken as,
\begin{equation}
\rho_n=\otimes_{i=1}^{N} \mathbb{I}/2.
\label{maximally_mixed}
\end{equation}
Because the Hamiltonian and the initial state of nuclear spins are both separable, the coherence of the electron spin is given by the multiplication of those from individual nuclear spins,
\begin{equation}
C = C^{(\text{b})} \prod_{i=1}^N C^{(i)},
\label{coherence_final}
\end{equation}
where $C^{(i)}$ is the decoherence from the $i$th nuclear spins, for an even number of $\pi$ pulses $N_\pi$,
\begin{equation}
\begin{split}
C^{(i)} & = 1 - 2 (1 - (n_z^{(i)})^2)\sin^2{\Big(\frac{\omega_0^{(i)}\tau}{4}\Big)}
 \sin^2{\Big(\frac{\omega_1^{(i)}\tau}{4}\Big)}
  \frac{\sin^2(N_\pi\phi^{(i)}/2)}{\cos^2(\phi^{(i)}/2)}, \\
\phi^{(i)} & = \arccos{\left(\cos{\Big(\frac{\omega_0^{(i)}\tau}{2}\Big)}\cos{\Big(\frac{\omega_1^{(i)}\tau}{2}\Big)}-n_z^{(i)}\sin{\Big(\frac{\omega_0^{(i)}\tau}{2}\Big)}\sin{\Big(\frac{\omega_1^{(i)}\tau}{2}\Big)}\right)},
\label{coherence_single}
\end{split}
\end{equation}
and $C^{(b)}$ is the decoherence from the residual spin bath that is taken as a classical time-varying noise,
\begin{equation}
C^{(\text{b})} = \exp{\left(-\frac{1}{2}\Big(\frac{1}{\pi} \sigma^{(\text{b})} N_\pi \tau\Big)^2\sum_{k=1}^\infty \frac{1}{(2k-1)^2}\left(\frac{\sin{[(\lvert\omega_{\text{L}}\rvert\tau-(2k-1)\pi)N_\pi/2]}}{(\lvert\omega_{\text{L}}\rvert\tau-(2k-1)\pi)N_\pi/2}\right)^2\right)},
\label{coherence_bath}
\end{equation}
where $\sigma^{(\text{b})}$ denotes the noise strength. Eq.~(\ref{coherence_final}) is utilized to fit the experimental results for extracting the hyperfine parameters of nuclear spins for two NV centers, as shown in Fig.~\ref{two-body_13C} and Fig.~\ref{three-body_13C}.

\section*{Polarization measurements for nuclear spins}
\begin{figure}
\centering
\begin{overpic}[width=1.0\textwidth]{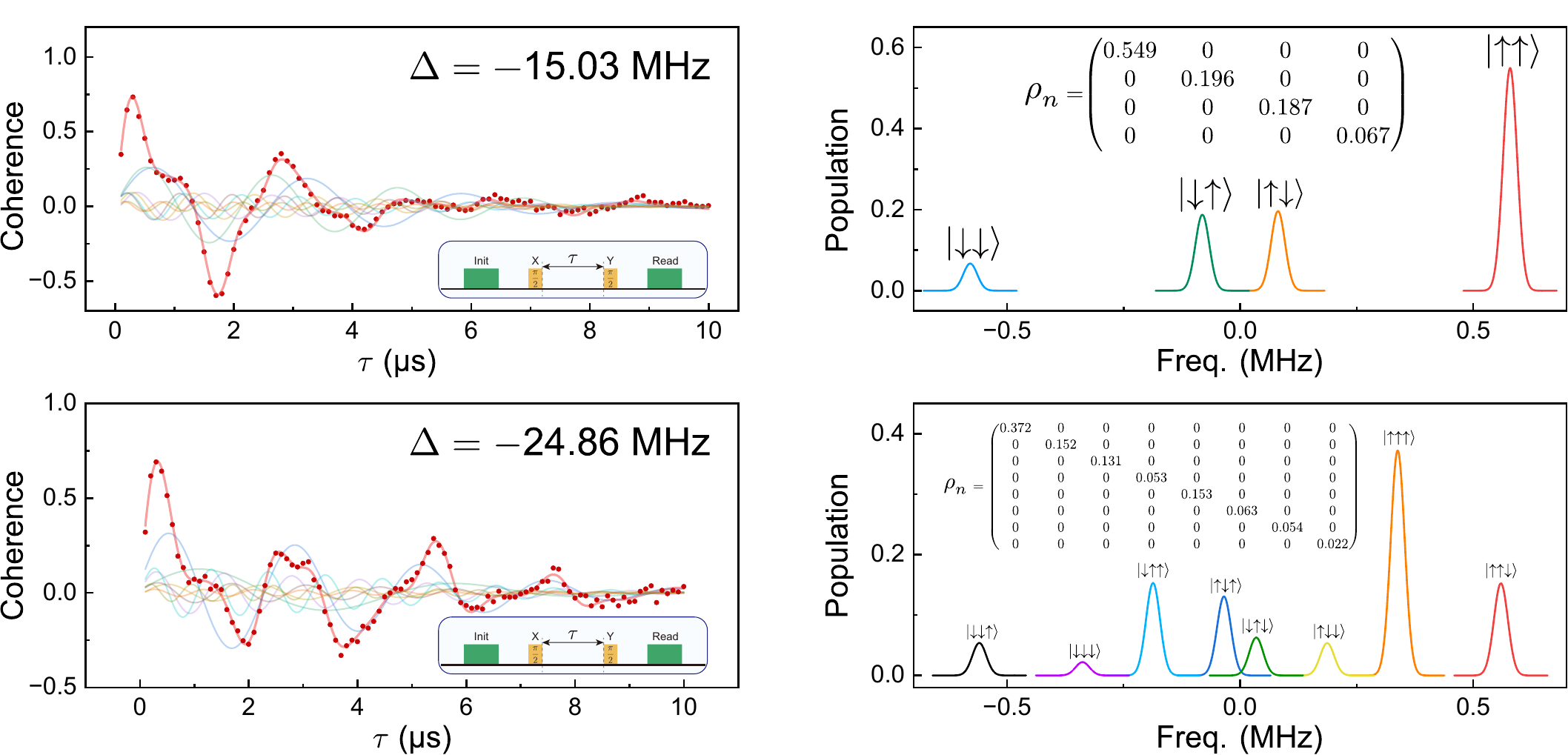}
	\put (-0.7, 46.5) {\figfont{(a)}}
	\put (52.1, 46.5) {\figfont{(b)}}
	\put (-0.7, 22.4) {\figfont{(c)}}
	\put (52.1, 22.4) {\figfont{(d)}}
\end{overpic}
\caption{Measurement of spin polarization after optical pumping. (a,c) The polarization measurement of nuclear spins by applying the Ramsey sequence for two NV centers under the energy gap $\Delta$ = $-$15.03~MHz and $-$24.86~MHz respectively. Eight light curves stand for eight components with different frequencies in Eq.~(\ref{coherence_Ramsey_i_f}). (b,d) The initial states of two and three nuclear spins for two NVs by fitting the results in (a,c).}\label{polarization}
\end{figure}

According to Eq.~(\ref{coherence_r}), it is necessary to measure the initial state of nuclear spins for explaining two-body and three-body dynamical results in theory. The Ramsey experiments are performed here to measure the initial state. The framework used for analyzing the DD sequences also works for the Ramsey case except
\begin{equation}
C_{\text{i}} = -\text{Im}[\text{Tr}(U_{0} \rho_n U_{1}^{\dag})],\text{ } U_{0} = u_{0}^2 = e^{-i H_0 \tau},\text{ } U_{1} = u_{1}^2 = e^{-i H_1 \tau},
\label{coherence_Ramsey_i}
\end{equation}
where the symbol `Im' means that the imaginary part of the coherence is read out since the second $\pi/2$ pulse of the Ramsey sequence in this work is applied along the y axis. For a single nuclear spin, the density matrix can be denoted as
\begin{equation}
\rho_n = \frac{\mathbb{I} + P\sigma_z}{2},
\label{density}
\end{equation}
where $P$ is the polarization of the spin. The Hamiltonian $H_0$ and $H_1$ in two subspaces are given just as above,
\begin{equation}
H_0 = \omega_0 I_z = \omega_\text{L} I_z,\text{ } H_1 = \omega_1 \mathbf{n}\cdot\mathbf{I} = (\omega_\text{L}+A_\parallel) I_z + A_\perp I_x.
\label{H0H1}
\end{equation}
After some algebraic calculations, the resultant coherence is written as
\begin{equation}
\begin{split}
C = & \left(\cos{\frac{\omega_0\tau}{2}}\cos{\frac{\omega_1\tau}{2}} + n_z\sin{\frac{\omega_0\tau}{2}} \sin{\frac{\omega_1\tau}{2}}\right) \\
& + iP\left(n_z\sin{\frac{\omega_1\tau}{2}}\cos{\frac{\omega_0\tau}{2}}-\sin{\frac{\omega_0\tau}{2}}\cos{\frac{\omega_1\tau}{2}}\right),
\label{coherence_Ramsey_single}
\end{split}
\end{equation}
Since $n_z$ is larger than 0.9 for the nuclear spins for two NV centers,  with ignoring high-frequency oscillating terms with small amplitudes $\frac{1-n_z}{2}$, Eq.~(\ref{coherence_Ramsey_single}) can be approximated to
\begin{equation}
C \approx \frac{1+n_z}{2}e^{i(\frac{\omega_1-\omega_0}{2})\tau}\left(\frac{1+P}{2}+\frac{1-P}{2}e^{-i(\omega_1-\omega_0)\tau}\right).
\label{coherence_Ramsey_single_2}
\end{equation}
Considering multiple independent nuclear spins, the final coherence is given by multiplication just as above,
\begin{equation}
C \approx A e^{-i\delta\tau} \prod_i \left(\frac{1+P^{(i)}}{2}+\frac{1-P^{(i)}}{2}e^{-i\delta^{(i)}\tau}\right).
\label{coherence_Ramsey_f}
\end{equation}
where $A = \prod_i \frac{1+n_z^{(i)}}{2}$, $\delta^{(i)} = \omega_1^{(i)}-\omega_0^{(i)}$, and the total $\delta$ contains all $\delta^{(i)}$ as well as the detuning used in the Ramsey experiments. There are three nuclear spins involved in both NVs, so the function used for fitting the Ramsey results is obtained below,
\begin{equation}
\begin{split}
C_{\text{i}} \approx & \frac{A}{8} e^{-(\tau/T_2^\ast)^2} \big\{(1+P^{(1)})(1+P^{(2)})(1+P^{(3)}) \sin{(\delta\tau+\varphi_0)}\\
& + (1-P^{(1)})(1+P^{(2)})(1+P^{(3)}) \sin{[(\delta+\delta^{(1)})\tau+\varphi_1]}\\
& + (1+P^{(1)})(1-P^{(2)})(1+P^{(3)}) \sin{[(\delta+\delta^{(2)})\tau+\varphi_2]}\\
& + (1+P^{(1)})(1+P^{(2)})(1-P^{(3)}) \sin{[(\delta+\delta^{(3)})\tau+\varphi_3]}\\
& + (1-P^{(1)})(1-P^{(2)})(1+P^{(3)}) \sin{[(\delta+\delta^{(1)}+\delta^{(2)})\tau+\varphi_4]}\\
& + (1-P^{(1)})(1+P^{(2)})(1-P^{(3)}) \sin{[(\delta+\delta^{(1)}+\delta^{(3)})\tau+\varphi_5]}\\
& + (1+P^{(1)})(1-P^{(2)})(1-P^{(3)}) \sin{[(\delta+\delta^{(2)}+\delta^{(3)})\tau+\varphi_6]}\\
&+ (1-P^{(1)})(1-P^{(2)})(1-P^{(3)}) \sin{[(\delta+\delta^{(1)}+\delta^{(2)}+\delta^{(3)})\tau+\varphi_7]}\big\},
\label{coherence_Ramsey_i_f}
\end{split}
\end{equation}
where the decoherence factor $e^{-(\tau/T_2^\ast)^2}$ and the initial phases incurred by the finite duration of $\pi/2$ pulses are included. As shown in Fig.~\ref{polarization}(a), the polarization parameters of the nuclear spins involved in two-body dynamics under the energy gap $\Delta$ = $-$15.03~MHz are obtained to be $P^{(1)}$ = 0.49(4) and $P^{(2)}$ = 0.47(4) with $P^{(3)}$ fixed to be zero for the third spin. The deviation of data from the fitting curve after 6~\textmu s might be originated from other weakly-coupled spins with the electron spin. Accordingly, the density matrix of the first two spins is obtained to be the input of two-body dynamics under the same gap, as displayed in Fig.~\ref{polarization}(b). Executing similar operations for the nuclear spins involved in three-body dynamics outputs the polarization parameters $P^{(1)}$ = 0.42(2), $P^{(2)}$ = 0.48(3) and $P^{(3)}$ = 0.42(2) in Fig.~\ref{polarization}(c) and the density matrix in Fig.~\ref{polarization}(d).

\section*{Approximate models and fitting}
With the derivations in the above sections, approximate theoretical models can be built here for fitting the dynamical results in the main text. First, both the direct dipolar interaction and the J-coupling in the subspace $\ket{m_S=+1}$ are too weak to take effect within the evolution duration in the experiment. The effective field in the subspace $\ket{m_S=0}$ induced by the non-secular terms of the hyperfine coupling tensor shifts the Larmor frequency, and thus must be included in the model. As for the J-coupling in the subspace $\ket{m_S=0}$, all terms that are non-commutative with the operator $I_z^{(1)}+I_z^{(2)}$ are abandoned in the model. Among them, two flip-flop terms play a crucial role in the spin dynamics, as shown in the main text, while the zz-component $A^{(12)}_{zz}$ slightly widens the single-spin dips within the evolution times of this work but is still included in fitting two-body dynamical results. With the analyses above, the Hamiltonian $H_0$ and $H_1$ of two nuclear spins for calculating the coherence of Eq.~(\ref{coherence_r}) are given by
\begin{equation}
\begin{split}
H_0 = & \sum_{i=1}^{2} (\omega_\text{L}+b_z^{(i)}) I_z^{(i)} + X^{(12)}\cos{\varphi^{(12)}}(I_x^{(1)}I_x^{(2)}+I_y^{(1)}I_y^{(2)}) \\
& + X^{(12)}\sin{\varphi^{(12)}}(-I_x^{(1)}I_y^{(2)}+I_y^{(1)}I_x^{(2)}) + A^{(12)}_{zz} I_z^{(1)}I_z^{(2)},
\label{H0_two_0}
\end{split}
\end{equation}
\begin{equation}
H_1 = \sum_{i=1}^{2} (\omega_\text{L}+A_\parallel^{(i)}) I_z^{(i)} + A_\perp^{(i)}(\cos{\varphi^{(i)}}I_x^{(i)} + \sin{\varphi^{(i)}}I_y^{(i)}),
\label{H1_two_0}
\end{equation}
where $b_z^{(i)}$ denotes the parallel component of the effective field in the subspace $\ket{m_S=0}$. The transverse components $b_x^{(i)}$ and $b_y^{(i)}$ are small enough compared with the Larmor frequency $\omega_\text{L}$, and thus are ignored in the model. Based on Eq.~(\ref{natural}), an approximate expression of two flip-flop terms $X^{(12)}\cos{\varphi^{(12)}}$ and $X^{(12)}\sin{\varphi^{(12)}}$ can be derived, and the latter is quite small so as to give a near-zero phase $\varphi^{(12)}$.

In fact, the phase parameters in Eq.~(\ref{H0_two_0}, \ref{H1_two_0}) are superfluous for fitting the dynamical results in the main text, because the relative angle $\varphi^{(i)}$ of the hyperfine terms $A_{zx}^{(i)}$ and $A_{zy}^{(i)}$ can be absorbed into the flip-flop phase,
\begin{equation}
\begin{split}
H_0 = & \sum_{i=1}^{2} (\omega_\text{L}+b_z^{(i)}) I_z^{(i)} + X^{(12)}\cos{\Phi^{(12)}}(I_x^{(1)}I_x^{(2)}+I_y^{(1)}I_y^{(2)}) \\
& + X^{(12)}\sin{\Phi^{(12)}}(-I_x^{(1)}I_y^{(2)}+I_y^{(1)}I_x^{(2)}) + A^{(12)}_{zz} I_z^{(1)}I_z^{(2)},
\label{H0_two}
\end{split}
\end{equation}
\begin{equation}
H_1 = \sum_{i=1}^{2} (\omega_\text{L}+A_\parallel^{(i)}) I_z^{(i)} + A_\perp^{(i)} I_x^{(i)},
\label{H1_two}
\end{equation}
\begin{equation}
\Phi^{(12)} = \varphi^{(12)}-\varphi^{(1)}+\varphi^{(2)} \approx \varphi^{(2)}-\varphi^{(1)}.
\label{chiral_angle}
\end{equation}
Considering that the phase $\varphi^{(12)}$ is near zero, $\Phi^{(12)}$ is approximately equal to the difference of two azimuth angles $\varphi^{(1)}$ and $\varphi^{(2)}$, which is the key to chiral discrimination. Now the model contains only five unknown parameters for fitting the dynamical results in Fig.~2 of the main text, and it is natural to extend the method above to multiple nuclear spins but there will be too many parameters involved for a great number of nuclear spins to be fitted ($\sim$$N^2$ parameters for $N$ spins). For three nuclear spins, it is still simple to fit the three-body dynamical results in Fig.~3 of the main text by using the Hamiltonian $H_0$ and $H_1$,
\begin{equation}
\begin{split}
H_0 = & \sum_{i=1}^{3} (\omega_\text{L}+b_z^{(i)}) I_z^{(i)} \\
& + X^{(12)}[\cos{\Phi^{(12)}}(I_x^{(1)}I_x^{(2)}+I_y^{(1)}I_y^{(2)}) + \sin{\Phi^{(12)}}(-I_x^{(1)}I_y^{(2)}+I_y^{(1)}I_x^{(2)})] \\
& + X^{(23)}[\cos{\Phi^{(23)}}(I_x^{(2)}I_x^{(3)}+I_y^{(2)}I_y^{(3)}) + \sin{\Phi^{(23)}}(-I_x^{(2)}I_y^{(3)}+I_y^{(2)}I_x^{(3)})] \\
& + X^{(31)}[\cos{\Phi^{(31)}}(I_x^{(3)}I_x^{(1)}+I_y^{(3)}I_y^{(1)}) + \sin{\Phi^{(31)}}(-I_x^{(3)}I_y^{(1)}+I_y^{(3)}I_x^{(1)})],
\label{H0_three}
\end{split}
\end{equation}
\begin{equation}
H_1 = \sum_{i=1}^{3} (\omega_\text{L}+A_\parallel^{(i)}) I_z^{(i)} + A_\perp^{(i)} I_x^{(i)},
\label{H1_three}
\end{equation}
where the chirality is manifested in the signs of the phases $\Phi^{(12)}$, $\Phi^{(23)}$ and $\Phi^{(31)}$. The zz-coupling terms $A^{(12)}_{zz}$, $A^{(23)}_{zz}$ and $A^{(31)}_{zz}$ are not included in Eq.~(\ref{H0_three}), since their effects are negligible in fitting three-body dynamical results under three energy gaps.

\begin{figure}
\centering
\begin{overpic}[width=1.0\textwidth]{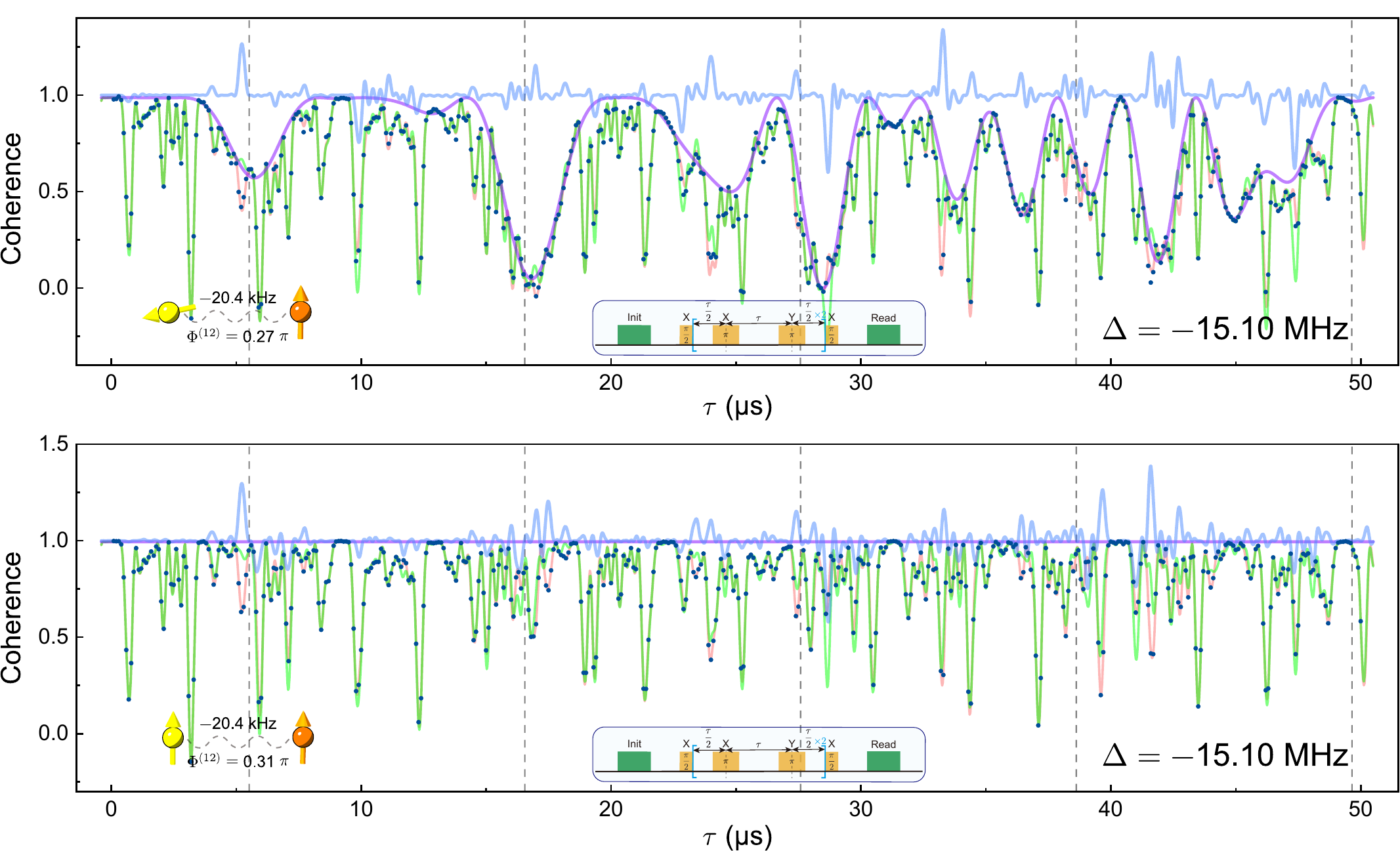}
	\put (-0.7, 59.7) {\figfont{(a)}}
	\put (-0.7, 29.1) {\figfont{(b)}}
\end{overpic}
\caption{Simulation of two-body spin dynamics and fitting with the two-body model under different spin polarization. (a) Simulation of the dynamics with one nuclear spin unpolarized and the other fully polarized under the energy gap $\Delta$ = $-$15.10~MHz by applying the XY4 sequence. The red line is the fitting curve with $\Phi^{(12)}$ = 0.27~$\pi$, while the green one is plotted with $\Phi^{(12)}$ = $-$0.27~$\pi$. The blue one is the difference of these two lines with an offset of 1, and the purple one depicts the coupling-induced envelope. (b) Simulation of the dynamics of two fully polarized nuclear spins. The red line is the fitting curve with $\Phi^{(12)}$ = 0.31~$\pi$ and the green one is plotted with $\Phi^{(12)}$ = $-$0.31~$\pi$.}\label{simulation_polarization}
\end{figure}

\begin{figure}
\centering
\begin{overpic}[width=1.0\textwidth]{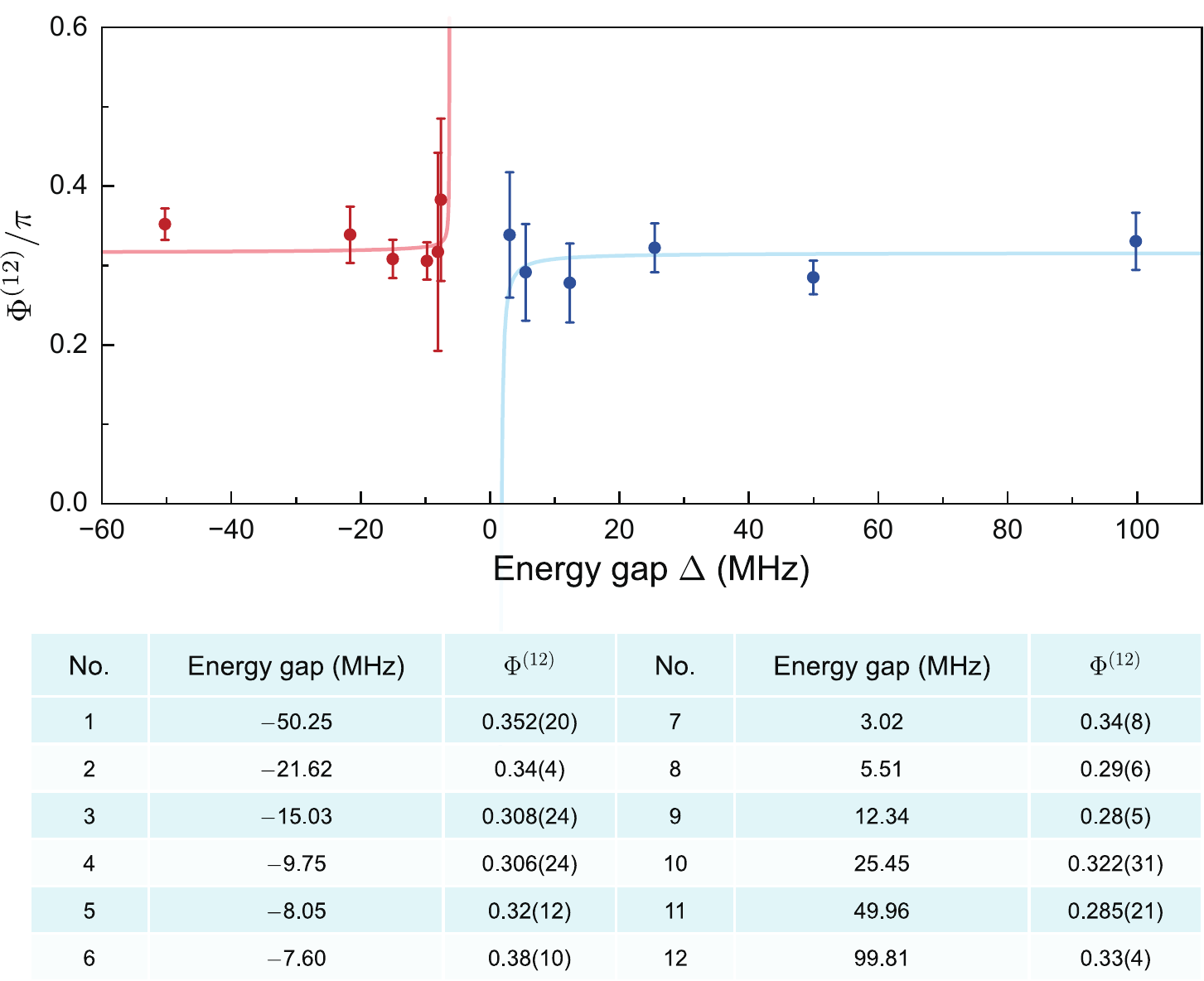}
	\put (-0.2, 80.2) {\figfont{(a)}}
	\put (-0.2, 29.0) {\figfont{(b)}}
\end{overpic}
\caption{Measurement of the chiral angle by recording two-body dynamics under different energy gaps. (a) The measured chiral angles under twelve energy gaps with all values collected in (b). The experimental results agree well with the perturbation results (solid lines) by using the hyperfine tensors from the DFT calculations. The asymptotic value under large gaps is equal to 0.316 $\pi$. The dynamical results under three gaps 99.81 MHz, 25.45 MHz and $-$15.03 MHz, are displayed in the main text for illustrating the process of symmetry breaking with respect to spin polarization.}\label{gaps}
\end{figure}

In order to verify whether the above models with the number of the parameters reduced are still precise enough, numerical simulations combined with fitting are performed here. As shown in Fig.~\ref{simulation_polarization}, two-body dynamical behaviors under  different spin polarization are numerically simulated by using the natural Hamiltonian with the hyperfine tensors from the calculations based on the density functional theory (DFT). It is difficult to derive an analytical formula of the coherence in Eq.~(\ref{coherence_r}) for the model described by Eq.~(\ref{H0_two}, \ref{H1_two}), so that it is not convenient to perform the fitting with commercial softwares. All fits in this work are manually programmed on our own by adopting the Gauss–Newton algorithm to minimize a squared residual error, and using the Jacobian matrix to calculate the errors of the fitting parameters combined with the minimum residual error and the correction factor from the $t$ distribution. It is clearly seen from Fig.~\ref{simulation_polarization} that the data points are quite close to the fitting curve compared with the experimental errors, signifying that the model of Eq.~(\ref{H0_two}, \ref{H1_two}) is precise enough to be used for fitting the experimental results. The purple envelopes in Fig.~\ref{simulation_polarization} as well as the figures of the main text, are calculated by modifying the Hamiltonian in the subspace $\ket{m_S=+1}$ as
\begin{equation}
H_1^\text{e} = \sum_{i} \omega_1^{(i)} I_z^{(i)},\text{ } \omega_1^{(i)} = -\sqrt{(\omega_\text{L}+A_\parallel^{(i)})^2+(A_\perp^{(i)})^2},
\label{envelope}
\end{equation}
with which sharp dips are removed for better depicting the coupling-induced dynamical behaviors. In Fig.~\ref{simulation_polarization}(a), one of two nuclear spins is fully polarized with the polarization $P=1$ and the other is unpolarized with $P=0$. Without polarizing both spins, the spin dynamics of two enantiomers have been evidently different in single-body sharp dips, just like Fig.~2(a) of the main text. It can be easily understood from the fact that a nonzero pseudoscalar $\hat{\mathbf{n}}^{\text{NV}}\cdot\mathbf{I}^{(i)}$ can now be constructed with the polarized spin to break the mirror symmetry or directly inferred from Eq.~(2) in the main text. Figure~\ref{simulation_polarization}(b) exhibits the two-body dynamics of two fully polarized nuclear spins. It is found that the modulation envelope caused by the flip-flop dynamics disappear in the final dynamical results, because for fully polarized nuclear spins, there is no population in the flip-flop states $\ket{\uparrow\downarrow}$ and $\ket{\downarrow\uparrow}$. Nevertheless, the spin dynamics of two enantiomers are still different in Fig.~\ref{simulation_polarization}(b) in that the azimuth angles of nuclear spins are still retained by the flip-flop term. In the experimetns, various two-body dynamics of two nuclear spins are recorded under twelve energy gaps. Based on the model above, chiral angles are extracted from the dynamical results by fitting them with the corresponding initial state and the hyperfine parameters input, with all values given in Fig.~\ref{gaps}. Among them, under large energy gaps (99.81 MHz, 49.96 MHz and $-$50.25 MHz in Fig.~\ref{gaps}(b)), both nuclear spins are almost unpolarized and take the maximum mixed state as the initial state in the fitting.

\section*{Symmetry transformations in spin space}
A detailed derivation of Eq.~(2) in the main text is given in this section. Considering the two-body model of Eq.~(\ref{H0_two}, \ref{H1_two}), the coherence of the electron spin under the DD sequences can be expressed as, based on Eq.~(\ref{coherence_r}),
\begin{equation}
C(\Phi^{(12)},A^{(12)}_{zz},P_1,P_2)  = \text{Tr}\big[U_0(\Phi^{(12)},A^{(12)}_{zz}) \rho_n(P_1,P_2) U_1^{\dag}(\Phi^{(12)},A^{(12)}_{zz})\big],
\label{coherence_appendix}
\end{equation}
as a function of the chiral angle $\Phi^{(12)}$ and spin polarization $P_1$ and $P_2$. As described in the main text, the operators of two nuclear spins for the mirror reflection, the time reversal and the joint transformation are given by $M=\sigma_y^{(1)} \otimes \sigma_y^{(2)}$, $T=\sigma_y^{(1)} \otimes \sigma_y^{(2)}\kappa$ and $MT=\kappa$, respectively, where $\kappa$ means to take the complex conjugate of the following expression. Under these transformations, the spin operators of two nuclear spins change in the following manner:
\begin{align}
I_x^{(i)}, I_y^{(i)}, I_z^{(i)} & \overset{M}{\longrightarrow} -I_x^{(i)}, I_y^{(i)}, -I_z^{(i)} \label{spin_M} \\
I_x^{(i)}, I_y^{(i)}, I_z^{(i)} & \overset{T}{\longrightarrow} -I_x^{(i)}, -I_y^{(i)}, -I_z^{(i)} \label{spin_T} \\
I_x^{(i)}, I_y^{(i)}, I_z^{(i)} & \overset{MT}{\longrightarrow} I_x^{(i)}, -I_y^{(i)}, I_z^{(i)} \label{spin_MT}
\end{align}
and behaves like angular momenta. The initial state of nuclear spins, $\rho_n(P_1,P_2)$, can be denoted as $\otimes_{i=1}^{2} (\mathbb{I}/2+P_i I_z^{(i)})$, and thus, it transforms as
\begin{equation}
\rho_n(P_1,P_2) \overset{M,T}{\longrightarrow} \rho_n(-P_1,-P_2),\text{ } \rho_n(P_1,P_2) \overset{MT}{\longrightarrow} \rho_n(P_1,P_2).\label{state_transformation}
\end{equation}
Next, the transformation relations for the Hamiltonian $H_0$ and $H_1$ denoted by Eq.~(\ref{H0_two}, \ref{H1_two}), are derived as follows:
\begin{align}
H_0(\Phi^{(12)},A^{(12)}_{zz}), H_1 & \overset{M}{\longrightarrow} -H_0(\pi-\Phi^{(12)},-A^{(12)}_{zz}), -H_1 \label{hamiltonian_M} \\
H_0(\Phi^{(12)},A^{(12)}_{zz}), H_1 & \overset{T}{\longrightarrow} -H_0(\pi+\Phi^{(12)},-A^{(12)}_{zz}), -H_1 \label{hamiltonian_T} \\
H_0(\Phi^{(12)},A^{(12)}_{zz}), H_1 & \overset{MT}{\longrightarrow} H_0(-\Phi^{(12)},A^{(12)}_{zz}), H_1 \label{hamiltonian_MT}
\end{align}
Using the relation $i \overset{T}{\rightarrow} -i$, the evolution operators $U_0(\Phi^{(12)})$ and $U_1(\Phi^{(12)})$ given by Eq.~(\ref{U0U1}-\ref{u0u1}) that are encoded into the spin coherence of Eq.~(\ref{coherence_appendix}), transform as
\begin{align}
U_0(\Phi^{(12)},A^{(12)}_{zz}), U_1(\Phi^{(12)},A^{(12)}_{zz}) & \overset{M}{\longrightarrow} U_0^{\dag}(\pi-\Phi^{(12)},-A^{(12)}_{zz}), U_1^{\dag}(\pi-\Phi^{(12)},-A^{(12)}_{zz}) \label{evolution_M} \\
U_0(\Phi^{(12)},A^{(12)}_{zz}), U_1(\Phi^{(12)},A^{(12)}_{zz}) & \overset{T}{\longrightarrow} U_0(\pi+\Phi^{(12)},-A^{(12)}_{zz}), U_1(\pi+\Phi^{(12)},-A^{(12)}_{zz}) \\
U_0(\Phi^{(12)},A^{(12)}_{zz}), U_1(\Phi^{(12)},A^{(12)}_{zz}) & \overset{MT}{\longrightarrow} U_0^{\dag}(-\Phi^{(12)},A^{(12)}_{zz}), U_1^{\dag}(-\Phi^{(12)},A^{(12)}_{zz}) \label{evolution_MT}
\end{align}
For the operation of the matrix trace in Eq.~(\ref{coherence_appendix}), under the mirror reflection and the time reversal, there exist the following relations, $\text{Tr}[O]=\text{Tr}[MOM^{-1}]$ and $\text{Tr}[O]=\text{Tr}[TOT^{-1}]^{\ast}$. Integrating all the above outputs the relations about the coherence of Eq.~(\ref{coherence_appendix}),
\begin{align}
C(\Phi^{(12)},A^{(12)}_{zz},P_1,P_2) & = \text{Tr}\big[U_0^{\dag}(\pi-\Phi^{(12)},-A^{(12)}_{zz}) \rho_n(-P_1,-P_2) U_1(\pi-\Phi^{(12)},-A^{(12)}_{zz})\big] \label{M-symmetry_appendix} \\
& = \text{Tr}\big[U_0(\pi+\Phi^{(12)},-A^{(12)}_{zz}) \rho_n(-P_1,-P_2) U_1^{\dag}(\pi+\Phi^{(12)},-A^{(12)}_{zz})\big]^{\ast} \label{T-symmetry_appendix} \\
& = \text{Tr}\big[U_0^{\dag}(-\Phi^{(12)},A^{(12)}_{zz}) \rho_n(P_1,P_2) U_1(-\Phi^{(12)},A^{(12)}_{zz})\big]^{\ast}. \label{MT-symmetry_appendix}
\end{align}
From the equations above, it is obvious that the spin dynamics will give the unique value of the angle $\Phi^{(12)}$ provided that one of the nuclear spins is polarized. When both spins are unpolarized with $P_1=P_2=0$ and $\rho_n(0,0)=\mathbb{I}_{4\times4}/4$, Eq.~(\ref{MT-symmetry_appendix}) take the form of
\begin{align}
C(\Phi^{(12)},A^{(12)}_{zz},0,0) & = \text{Tr}\big[U_0^{\dag}(-\Phi^{(12)},A^{(12)}_{zz}) U_1(-\Phi^{(12)},A^{(12)}_{zz})\big]^{\ast}/4 = C(-\Phi^{(12)},A^{(12)}_{zz},0,0), \label{MT-symmetry_unpolarized}
\end{align}
which explains the two-fold degeneracy in the real and imaginary parts of the coherence in Fig.~1(f) of the main text. It is quite straightforward to generalize the above analysis and derivation on the symmetry transformations to the case of many-body dynamics of multiple nuclear spins.

\section*{Magnetic-field-inversion symmetry breaking}
In this section, we will give a proof on why the chiral angle $\Phi^{(12)}$ changes the sign after the magnetic field reversal. By combining Eq.~(\ref{Hamiltonian_0}) and Eq.~(\ref{hyperfine}), the whole Hamiltonian of the NV electron spin coupled to two $^{\text{13}}$C nuclear spins, as a function of the magnetic field $\mathbf{B}_0$ (along the NV axis), is rewritten as 
\begin{equation}
H = DS_z^2 - \gamma_e B_0 S_z - \gamma_n B_0(I_z^{(1)}+I_z^{(2)}) + \sum_{i=1}^2 \mathbf{S}\cdot\mathbf{A}^{(i)}\cdot\mathbf{I}^{(i)},
\label{Hamiltonian_whole}
\end{equation}
where $\omega_e$ and $\omega_\text{L}$ in Eq.~(\ref{Hamiltonian_0}) are expressed as $-\gamma_e B_0$ and $-\gamma_n B_0$ by taking the direction of $\mathbf{B}_0$ as the positive z-axis of the coordinate frame. The joint state of the electron spin and the nuclear spins after optical pumping can be given by
\begin{equation}
\rho_0 = \ket{m_S=0}\bra{m_S=0}_e \otimes_{i=1}^{2} (p_i\ket{\uparrow}\bra{\uparrow}_i+(1-p_i)\ket{\downarrow}\bra{\downarrow}_i).
\label{joint_state}
\end{equation}
After the magnetic field is reversed, under the original frame, $\mathbf{B}_0$ changes the sign and the hyperfine tensor $\mathbf{A}^{(i)}$ remains unchanged in the Hamiltonian
\begin{equation}
H' = DS_z^2 + \gamma_e B_0 S_z + \gamma_n B_0(I_z^{(1)}+I_z^{(2)}) + \sum_{i=1}^2 \mathbf{S}\cdot\mathbf{A}^{(i)}\cdot\mathbf{I}^{(i)},
\label{Hamiltonian_whole_reversal}
\end{equation}
and the direction of spin polarization follows the magnetic field, modifying the initial state as
\begin{equation}
\rho'_0 = \ket{m_S=0}\bra{m_S=0}_e \otimes_{i=1}^{2} ((1-p_i)\ket{\uparrow}\bra{\uparrow}_i + p_i\ket{\downarrow}\bra{\downarrow}_i).
\label{joint_state_reversal}
\end{equation}
In order to clearly show the difference in physics between Eq.~(\ref{Hamiltonian_whole_reversal}, \ref{joint_state_reversal}) and Eq.~(\ref{Hamiltonian_whole}, \ref{joint_state}), redefine the eigenstates of the electron spin and the nuclear spins as 
\begin{equation}
\begin{split}
\ket{m_S=+1} \rightarrow \ket{m_S=-1}&, \text{ } \ket{m_S=-1} \rightarrow \ket{m_S=+1} \\
\text{ } \ket{\uparrow} \rightarrow \ket{\downarrow}&, \text{ } \ket{\downarrow} \rightarrow \ket{\uparrow}
\label{state_redefinition}
\end{split}
\end{equation}
Under the new representation, the spin operators $S_{y/z}$ and $I_{y/z}$ are replaced by $-S_{y/z}$ and $-I_{y/z}$. The initial state is formally recovered to Eq.~(\ref{joint_state}), and the Hamiltonian transforms into
\begin{equation}
H' = DS_z^2 - \gamma_e B_0 S_z-\gamma_n B_0(I_z^{(1)}+I_z^{(2)}) + \sum_{i=1}^2 \mathbf{S}\cdot\mathbf{A'}^{(i)}\cdot\mathbf{I}^{(i)},
\label{Hamiltonian_whole_reversal_redefinition}
\end{equation}
where the hyperfine tensor $\mathbf{A'}^{(i)}$ can be explicitly represented as
\begin{equation}
\mathbf{A'}^{(i)} =
\begin{pmatrix}
A^{(i)}_{xx} & -A^{(i)}_{xy} & -A^{(i)}_{xz} \\
-A^{(i)}_{yx} & A^{(i)}_{yy} & A^{(i)}_{yz} \\
-A^{(i)}_{zx} & A^{(i)}_{zy} & A^{(i)}_{zz} \\
\end{pmatrix}.
\label{coupling_tensor_redefinition}
\end{equation}
Notice that $A'^{(i)}_{xy/yx}=-A^{(i)}_{xy/yx}$ and $A'^{(i)}_{xz/zx}=-A^{(i)}_{xz/zx}$, and the remaining elements remain unchanged. The change of the hyperfine tensors due to the field reversal can be equivalently obtained by still taking the direction of the magnetic field as the positive z-axis and keeping the x-axis unchanged after the field reversal. Under this new coordinate frame, the hyperfine tensor $\mathbf{A'}^{(i)}$ will directly take the same form as Eq.~(\ref{coupling_tensor_redefinition}). After the field reversal, it can be easily derived from Eq.~(\ref{coupling_tensor_redefinition}) that the azimuth angles $\varphi'^{(1)}$ and $\varphi'^{(2)}$ in Eq.~(\ref{H1_two_0}) change into $\pi-\varphi^{(1)}$ and $\pi-\varphi^{(2)}$, and the flip-flop phase $\varphi'^{(12)}$ in Eq.~(\ref{H0_two_0}) becomes the opposite $-\varphi^{(12)}$ based on Eq.~(\ref{natural}). Therefore, the chiral angle $\Phi'^{(12)}$, defined by Eq.~(\ref{chiral_angle}), has the relation $\Phi'^{(12)} = \varphi'^{(12)}-\varphi'^{(1)}+\varphi'^{(2)} = -\varphi^{(12)}+\varphi^{(1)}-\varphi^{(2)} = -\Phi^{(12)}$. Since the model of Eq.~(\ref{H0_two}, \ref{H1_two}) is utilized constantly to fit the dynamical results, the chiral angle $\Phi^{(12)}$ should change the sign after the field reversal, as validated experimentally in Fig.~2 of the main text.

\section*{Determination of the absolute configuration}
As described above, the sign of the chiral angle $\Phi^{(12)}$ obtained by fitting the dynamical results is dependent on the direction of the magnetic field. Though two enantiomers diverge in the spin dynamics under spin polarization, the absolute configuration is still undetermined. From the perspective of the pseudoscalar, the configuration determination of a chiral molecule demands finding a macroscopic polar vector to be correlated with a certain axis of the molecule being studied. For the case in this work, it is not difficult to determine the direction of the NV axis by applying an electric field. The electric field will shift the zero-field splitting (ZFS) of the electron spin, denoted by $D$ in Eq.~(\ref{Hamiltonian_0}), with a coefficient of $dD/dE$ = 0.35(2)~Hz$\cdot$cm/V~\cite{van1990electric}. If the electric field is applied along the direction of $\hat{\mathbf{n}}^{\text{NV}}$ (pointing from the vacancy to the nitrogen atom), the coefficient $dD/dE$ is calculated to be $-$0.37~Hz$\cdot$cm/V with the DFT, which agrees well with the measured value above. As such, the direction of the NV axis can be determined by measuring the slope of the ZFS as a function of the electric field.

\begin{figure}
\centering
\begin{overpic}[width=1.0\textwidth]{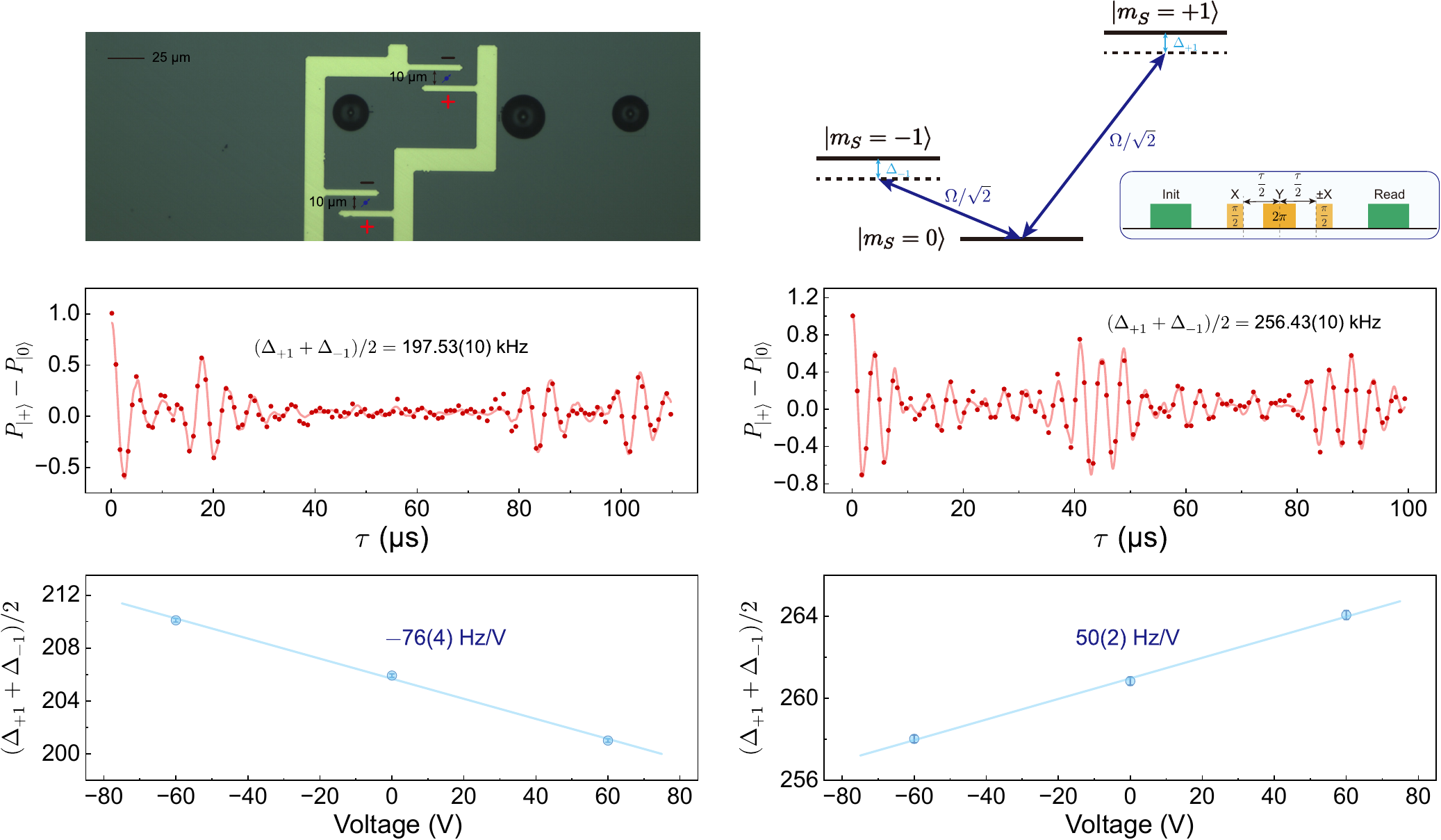}
	\put (-0.6, 56.1) {\figfont{(a)}}
	\put (51.2, 56.1) {\figfont{(b)}}
	\put (-0.6, 38.4) {\figfont{(c)}}
	\put (51.2, 38.4) {\figfont{(d)}}
	\put (-0.6, 18.4) {\figfont{(e)}}
	\put (51.2, 18.4) {\figfont{(f)}}
\end{overpic}
\caption{Determination of the direction of the NV axis with a static electric field. (a) Microscope image of the electrodes coated on the diamond surface for generating the electric field. (b) Microwave manipulation for two resonances of $\ket{0} \leftrightarrow \ket{+1}$ and $\ket{0} \leftrightarrow \ket{-1}$ and the pulse sequence for measuring the deviation from the ZFS. (c,d) The population difference between $\ket{+}$ and $\ket{0}$ measured for the NV center in Fig.~2 and Fig.~3 as a function of the interval $\tau$ under the sequence in (b) and the energy gap $\Delta$ $\approx$ $-$715~MHz. The red line in (c,d) is the fitting curve of Eq.~(\ref{ppopulation_difference_triplet_final}) with $(\Delta_{+1}+\Delta_{-1})/2$ = 197.53(10)~kHz (256.43(10)~kHz). (e,f) The measured deviation from the ZFS for the NV in (c,d) by varying the voltage applied to the electrodes.}\label{NV_axis}
\end{figure}

In order to measure the ZFS with a high enough precision, the method of the previous work~\cite{kucsko2013nanometre} is adopted here to resist the magnetic fluctuation by simultaneously controlling two transitions $\ket{m_S=0} \leftrightarrow \ket{m_S=+1}$ and $\ket{m_S=0} \leftrightarrow \ket{m_S=-1}$, as shown in Fig.~\ref{NV_axis}(b). The Hamiltonian for describing the microwave control after the rotating-wave approximation in the interaction picture can be written as
\begin{equation}
H_\text{c} = \Omega_{+1}/2 (\ket{+1}\bra{0} + \ket{0}\bra{+1}) + \Omega_{-1}/2 (\ket{-1}\bra{0} + \ket{0}\bra{-1}),
\label{control_triplet_0}
\end{equation}
where $\Omega_{+1}$ and $\Omega_{-1}$ are the control amplitudes of two transitions. With $\Omega_{+1}$ and $\Omega_{-1}$ assigned the same, two resonances can be synthesized as a whole,
\begin{equation}
H_\text{c} = \Omega/2 S_x = \Omega/2 (\ket{+}\bra{0} + \ket{0}\bra{+}),
\label{control_triplet}
\end{equation}
where the relation holds $\Omega/\sqrt{2} = \Omega_{+1} = \Omega_{-1}$. Equation~(\ref{control_triplet}) realizes the transition between $\ket{m_S=0}$ and $\ket{m_S=+}=(\ket{m_S=+1}+\ket{m_S=-1})/\sqrt{2}$, and leaves alone the third state $\ket{m_S=-}=(\ket{m_S=+1}-\ket{m_S=-1})/\sqrt{2}$. Next, we employ the pulse sequence~\cite{kucsko2013nanometre} in Fig.~\ref{NV_axis}(b) to measure the deviation from the ZFS $D$. The $\pi/2$ pulse prepares a superposition state of $\ket{m_S=0}$ and $\ket{m_S=+}$, while the $2\pi$ pulse in the middle swaps $\ket{m_S=+1}$ and $\ket{m_S=-1}$ so that the magnetic fluctuation is cancelled out. Finally, the sequence measures the difference between the average of two microwave frequencies and the ZFS, just like the Ramsey sequence.

However, some extra complexity is incurred by the existence of multiple strongly-coupled $^{\text{13}}$C nuclear spins. The overall Hamiltonian of the electron spin and the nearby nuclear spins during the free evolution can be represented as
\begin{equation}
H = \ket{0}\bra{0} \otimes H_{0} + \ket{+1}\bra{+1} \otimes H_{+1} + \ket{-1}\bra{-1} \otimes H_{-1},
\label{Hamiltonian_triplet}
\end{equation}
where $H_0$, $H_{+1}$ and $H_{-1}$ are the Hamiltonians of the nuclear spins in each subspace of the electron spin,
\begin{equation}
\begin{split}
H_0 & = \sum_{i=1}^{N} \omega_\text{L} I_z^{(i)} = \sum_{i=1}^{N} \omega_0^{(i)} I_z^{(i)}, \\
H_{+1} = \Delta_{+1} + \sum_{i=1}^{N} (\omega_\text{L} + & A_\parallel^{(i)}) I_z^{(i)} + A_\perp^{(i)}I_x^{(i)} = \Delta_{+1} + \sum_{i=1}^{N} \omega_{+1}^{(i)} \mathbf{n}_{+1}^{(i)} \cdot \mathbf{I}^{(i)}, \\
H_{-1} = \Delta_{-1} + \sum_{i=1}^{N} (\omega_\text{L} - & A_\parallel^{(i)}) I_z^{(i)} - A_\perp^{(i)}I_x^{(i)} = \Delta_{-1} + \sum_{i=1}^{N} \omega_{-1}^{(i)} \mathbf{n}_{-1}^{(i)} \cdot \mathbf{I}^{(i)},
\end{split}
\label{H0H1Hn1}
\end{equation}
where $\Delta_{+1}$ ($\Delta_{-1}$) is the detuning between the energy gap of $\ket{m_S=0} \leftrightarrow \ket{m_S=+1}$ ($\ket{m_S=-1}$) and the corresponding microwave frequency. The average of two detunings $(\Delta_{+1}+\Delta_{-1})/2$ measures the deviation from the ZFS, while the difference $(\Delta_{+1}-\Delta_{-1})$ is the gap between $\ket{m_S=+1}$ and $\ket{m_S=-1}$, which will be eliminated by the $2\pi$ pulse in the sequence. With the Hamiltonian given above, the state evolution can be calculated,
\begin{equation}
U = \ket{0}\bra{0} \otimes U_{0} + \ket{+1}\bra{+1} \otimes U_{+1} + \ket{-1}\bra{-1} \otimes U_{-1},
\label{unitary_triplet}
\end{equation}
where $U_0$, $U_{+1}$ and $U_{-1}$ are given by
\begin{equation}
\begin{split}
U_0 & = u_0 u_0 = e^{-i H_0 \tau}, \\
U_{+1} & = u_{-1} u_{+1} = e^{-i H_{-1} \tau/2} e^{-i H_{+1} \tau/2}, \\
U_{-1} & = u_{+1} u_{-1} = e^{-i H_{-1} \tau} e^{-i H_{+1} \tau/2}.
\end{split}
\label{U0U1Un1}
\end{equation}
Using some algebraic calculations, the populations of three states after the final $\pi/2$ pulse are obtained as follows,
\begin{align}
P_{\ket{0}} & = \frac{3}{8} + \frac{1}{16}\text{Re}[\text{Tr}(U_{+1}U_{-1}^{\dag})] - \frac{1}{8}\text{Re}[\text{Tr}(U_{+1}U_{0}^{\dag}+U_{-1}U_{0}^{\dag})], \\
P_{\ket{+}} & = \frac{3}{8} + \frac{1}{16}\text{Re}[\text{Tr}(U_{+1}U_{-1}^{\dag})] + \frac{1}{8}\text{Re}[\text{Tr}(U_{+1}U_{0}^{\dag}+U_{-1}U_{0}^{\dag})], \\
P_{\ket{-}} & = \frac{1}{4} - \frac{1}{8}\text{Re}[\text{Tr}(U_{+1}U_{-1}^{\dag})].
\label{population_triplet}
\end{align}
The populations of $\ket{m_S=0}$ and $\ket{m_S=+}$ can be exchanged by applying another $\pi$ pulse, and thus, the population difference of $\ket{m_S=0}$ and $\ket{m_S=+}$ can be obtained in the experiment,
\begin{equation}
P_{\ket{+}} - P_{\ket{0}} = \frac{1}{4}\text{Re}[\text{Tr}(U_{+1}U_{0}^{\dag}+U_{-1}U_{0}^{\dag})].
\label{population_difference_triplet}
\end{equation}
Substituting the Hamiltonian of Eq.~(\ref{H0H1Hn1}) into the unitary evolution of Eq.~(\ref{U0U1Un1}) outputs the expression for fitting the experimental results in Fig.~\ref{NV_axis}(c),
\begin{equation}
P_{\ket{+}} - P_{\ket{0}} = \cos{\Big(\frac{\Delta_{+1}+\Delta_{-1}}{2}\tau\Big)} \cdot \prod_{i=1}^N M^{(i)},
\label{ppopulation_difference_triplet_final}
\end{equation}
\begin{equation}
\begin{split}
M^{(i)} = & \cos{\Big(\frac{\omega_0^{(i)}\tau}{2}\Big)} \Bigg(\cos{\Big(\frac{\omega_{+1}^{(i)}\tau}{4}\Big)}\cos{\Big(\frac{\omega_{-1}^{(i)}\tau}{4}\Big)} - \mathbf{n}_{+1}^{(i)} \cdot \mathbf{n}_{-1}^{(i)} \sin{\Big(\frac{\omega_{+1}^{(i)}\tau}{4}\Big)} \\
& \sin{\Big(\frac{\omega_{-1}^{(i)}\tau}{4}\Big)}\Bigg)
+ \sin{\Big(\frac{\omega_0^{(i)}\tau}{2}\Big)} \Bigg(n_{+1,z}^{(i)}\sin{\Big(\frac{\omega_{+1}^{(i)}\tau}{4}\Big)}\cos{\Big(\frac{\omega_{-1}^{(i)}\tau}{4}\Big)} \\
& + n_{-1,z}^{(i)} \cos{\Big(\frac{\omega_{+1}^{(i)}\tau}{4}\Big)}\sin{\Big(\frac{\omega_{-1}^{(i)}\tau}{4}\Big)}\Bigg),
\end{split}
\end{equation}
where the detuning term $(\Delta_{+1}+\Delta_{-1})/2$ determines the deviation from the ZFS and $n_{+1,z}^{(i)}$ ($n_{-1,z}^{(i)}$) is the third component of $\mathbf{n}_{+1}^{(i)}$ ($\mathbf{n}_{-1}^{(i)}$).

\begin{figure}
\centering
\begin{overpic}[width=1.0\textwidth]{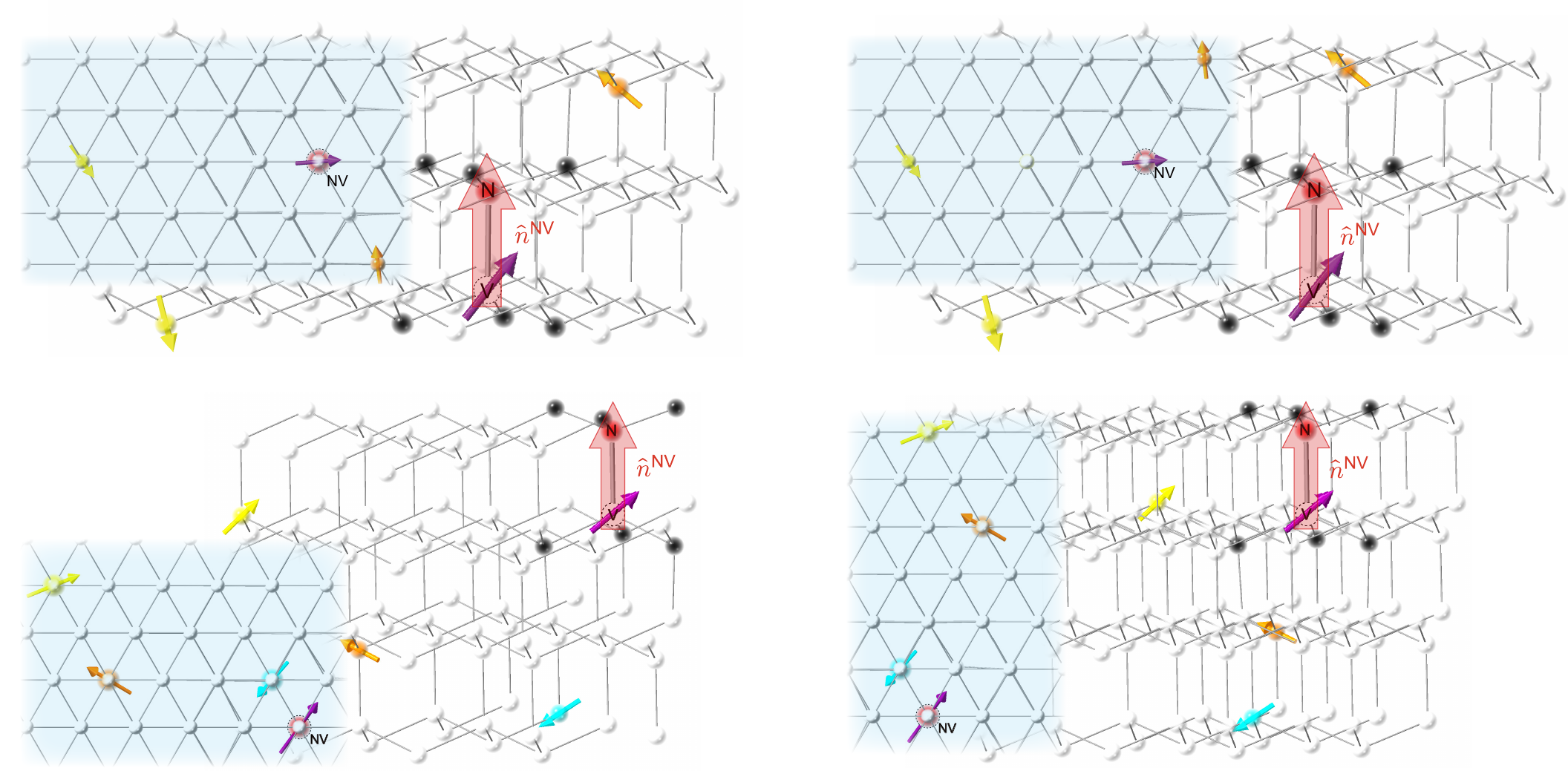}
	\put (-0.7, 47.4) {\figfont{(a)}}
	\put (52.1, 47.4) {\figfont{(b)}}
	\put (-0.7, 23.5) {\figfont{(c)}}
	\put (52.1, 23.5) {\figfont{(d)}}
\end{overpic}
\caption{Spatial configuration in diamond lattice. (a,b) The configurations of two enantiomers with two $^{\text{13}}$C nuclear spins involved for chiral discrimination. The inset is the top view of the lattice, from which it is clearly seen that two configurations are mirror-symmetric with each other. The chiral angle $\Phi^{(12)}$ in Fig.~2 of the main text determines the correct configuration to be (a). (c,d) The configurations of two enantiomers with three $^{\text{13}}$C spins involved for chiral discrimination. The chiral angles $\Phi^{(12)}$, $\Phi^{(23)}$ and $\Phi^{(31)}$ obtained from Fig.~3 of the main text determine the correct configuration to be (d).}\label{spatial_configuration}
\end{figure}

\begin{figure}
\centering
\begin{overpic}[width=1.0\textwidth]{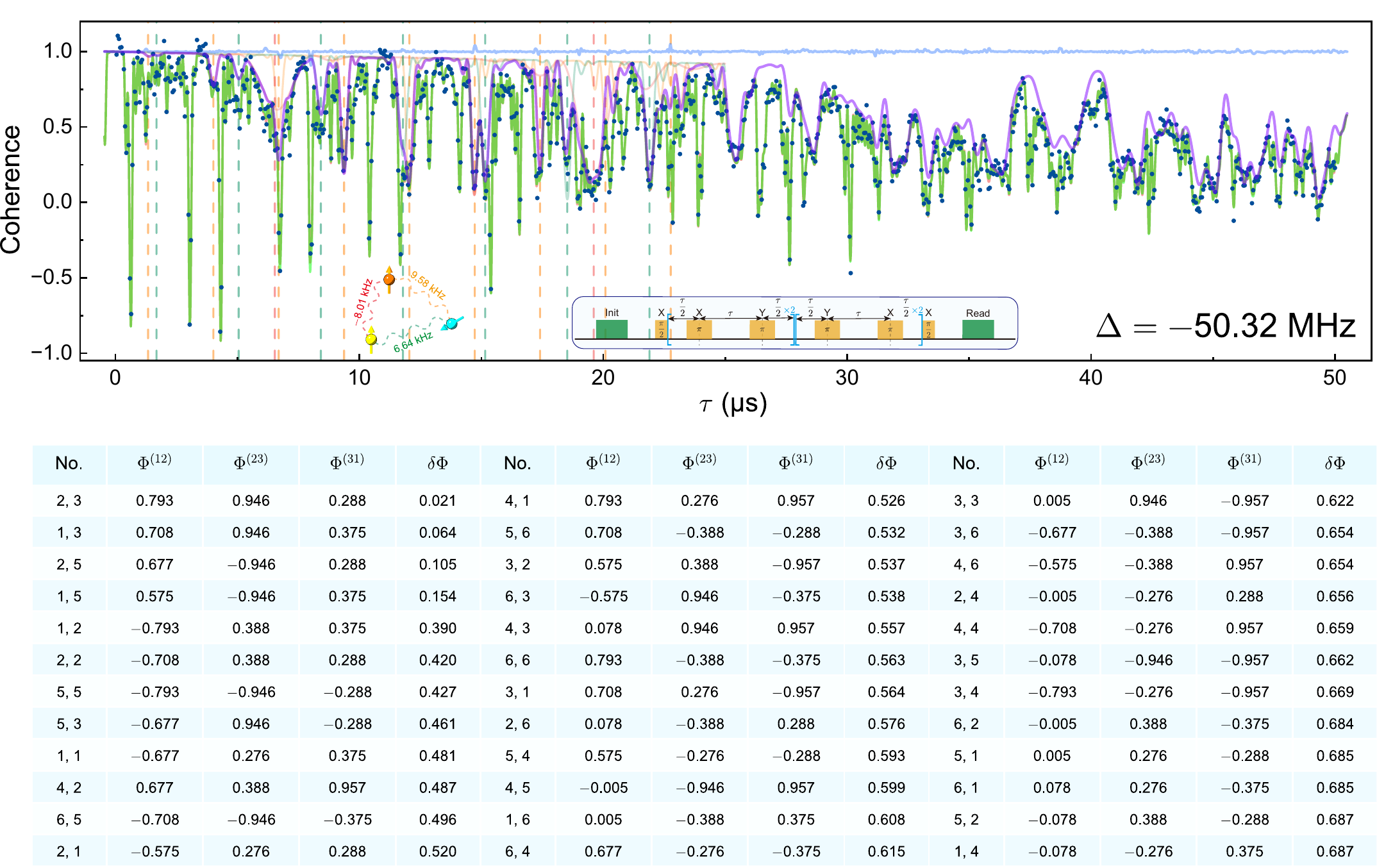}
	\put (-0.7, 61.7) {\figfont{(a)}}
	\put (-0.7, 30.8) {\figfont{(b)}}
\end{overpic}
\caption{Quantum three-body dynamics and configuration determination. (a) Quantum three-body dynamics of three nuclear spins under the energy gap $\Delta$ = $-$50.32~MHz by applying the XY8 sequence. (b) The DFT-based calculations of three chiral phases $\Phi^{(ij)}$ for thirty-six possible configurations of three nuclear spins. The configuration number is indicated by the positions of the first two spins. The unit of the phases $\Phi^{(ij)}$ is $\pi$, and $\delta\Phi$ is the root mean square of the difference between the calculated values and the experimental ones obtained by fitting the results in (a). Considering the periodicity of the phase, the operation of `modulo $2$ with an offset of $-1$' is executed in calculating the difference.}\label{possible_configurations_three}
\end{figure}

For applying an electric field to the NV centers in Fig.~2 and Fig.~3, a pattern of multiple electrodes is coated on the diamond surface, as depicted in Fig.~\ref{NV_axis}(a). The population difference $P_{\ket{+}}-P_{\ket{0}}$ is measured for both NVs by implementing the pulse sequence described above (Fig.~\ref{NV_axis}(b)), with the results displayed in Fig.~\ref{NV_axis}(c,d). Fitting the results with Eq.~(\ref{ppopulation_difference_triplet_final}), wherein the hyperfine parameters in Fig.~2 and Fig.~3 are employed, gives the detuning $(\Delta_{+1}+\Delta_{-1})/2$ = 197.53(10)~kHz in Fig.~\ref{NV_axis}(c) and 256.43(10)~kHz in Fig.~\ref{NV_axis}(d). Varying the voltage between two electrodes shifts the ZFS, and not two many points are recorded in Fig.~\ref{NV_axis}(e,f) given that the ZFS is affected by the variation of the local field. Based on the analysis combining the coefficient $dD/dE$ and the pattern in Fig.~\ref{NV_axis}(a), the sign of the slope in Fig.~\ref{NV_axis}(e) determines the absolute configuration of the first NV center to be the one in Fig.~\ref{spatial_configuration}(a), as well as the relative direction with respect to the magnetic field in the inset of Fig.~2 of the main text. Likewise, it is inferred from the fitting line in Fig.~\ref{NV_axis}(f), combined with the DFT calculations and the results in Fig.~\ref{possible_configurations_three}, that the second NV center has the configuration in Fig.~\ref{spatial_configuration}(d).

\section*{The DFT calculations}
The density functional theory (DFT) calculations are performed using the projector augmented-wave method~\cite{blochl1994projector} and implemented in the Vienna \textit{ab initio} simulation package (VASP)~\cite{kresse1993ab, kresse1994ab}. The NV$^-$ center is modeled in a 1728-atom diamond supercell (6$\times$6$\times$6 unit cells) with the optimized lattice parameter 3.567~\AA. Geometric relaxations are conducted using the Perdew-Burke-Ernzerhof exchange-correlation functional~\cite{perdew2008restoring} until residual forces on all atoms fall below 10$^{-\text{3}}$~eV/\AA. For the electronic structure calculations and wavefunction analysis, the Heyd-Scuseria-Ernzerhof (HSE06) functional~\cite{heyd2003hybrid} is employed with a plane-wave energy cutoff of 500~eV and $\Gamma$-point sampling of the Brillouin zone. The self-consistent field iterations utilize stringent convergence criteria with an energy tolerance of 10$^{-\text{6}}$~eV. To address long-range hyperfine interactions, a real-space integration scheme~\cite{takacs2024accurate} processing VASP outputs is implemented combined with the full spin density distribution on a fine real-space grid with a resolution of 0.036~\AA, enabling accurate computation of hyperfine tensors for distant $^{\text{13}}$C nuclei beyond conventional supercell limitations.

The ZFS tensor of the NV center under an external electric field is calculated using a model of NV-centered and hydrogen-terminated diamond lattice with up to 106 atoms. The DFT calculations are performed with the PBE0 functional with the def2-SVP basis set as implemented in the ORCA code~\cite{neeseORCAProgramSystem2012}. Tight convergence criteria are enforced for the calculations of geometry optimization and self-consistent field. The ZFS tensor incorporates the contributions from spin-spin interaction and spin-orbit coupling, wherein the spin-spin term is derived from the restricted spin density of singly occupied unrestricted natural orbitals, and the spin-orbit coupling is treated via the coupled-perturbed approach. The coefficient of $dD/dE$ = $-$0.37~Hz$\cdot$cm/V used in the experiment is obtained by applying an electric field along the direction of $\hat{\mathbf{n}}^{\text{NV}}$. When considering the structural relaxation induced by the electric field, the terminating hydrogen atoms are constrained to fixed positions, mimicking a bulk-like environment. The geometry optimization and the electronic structure are calculated under these constraints, which ensures the systematic evaluation of the electric effect on the ZFS tensor.

\clearpage
\bibliography{references.bib}
\clearpage

